\documentclass[a4paper,11pt]{article}
\usepackage{amsmath}
\usepackage{amsfonts}
\usepackage{amssymb}
\usepackage{enumerate}
\usepackage{verbatim}
\usepackage{graphicx}
\usepackage{float}
\usepackage{color}
\usepackage{tikz}
\usepackage{tikzscale}
\usepackage{afterpage}
\usepackage{pgfplots}
\usepackage{tabularx}
\usepackage{xcolor}
\usetikzlibrary{arrows, automata, shapes, backgrounds, calc, arrows.meta, tikzmark, positioning, decorations.pathmorphing}
\usepackage{longtable}
\usepackage{ragged2e}
\usepackage{natbib}
\usetikzlibrary{arrows, positioning}
\usepackage{rotating}
\usetikzlibrary{shapes.geometric, arrows}
\usepackage[hidelinks]{hyperref}
\usepackage[top=2cm,bottom=2cm,right=2cm,left=2cm]{geometry}
\usepackage[labelfont=bf]{caption} 
\usepackage{subcaption}
\usepackage[linesnumbered]{algorithm2e}
\usepackage{float}
\usepackage{booktabs}
\usepackage{longtable}
\usepackage{pdflscape}

\usepackage[symbol]{footmisc}

\title{\textbf{Identifying Important Inputs for Black Box Machine Learning Methods}}
\author{Mohammad Kaviul Anam Khan  \\ \textsc{email:} \textcolor{blue}{\href{mailto:kaviul.khan@mail.utoronto.ca}{kaviul.khan@mail.utoronto.ca}}}

\newcommand{\PO}[2]{#1^{(#2)}}

\newtheorem{theorem}{Theorem}

\begin{document}

%

\begin{center}
	\Large
	A Generalized Variable Importance Metric and it's Estimator for Black Box Machine Learning Models
	
	\small
	\vspace{0.4cm}
	Mohammad Kaviul Anam Khan\footnote[1]{Dalla Lana School of Public Health University ot Toronto. Email: kaviul.khan@mail.utoronto.ca}, Olli Saarela\footnote[1]{Dalla Lana School of Public Health University ot Toronto. Email: olli.saarela@utoronto.ca} and Rafal Kustra\footnote[1]{Dalla Lana School of Public Health University ot Toronto. Email: r.kustra@utoronto.ca} \\
	December 23rd, 2023
	
	\vspace{0.9cm}

\end{center}

\begin{abstract}
	In this paper we define a population parameter, ``Generalized Variable Importance Metric (GVIM)'', to measure importance of predictors for black box machine learning methods, where the importance is not represented by model-based parameter. GVIM is defined for each input variable, using the true conditional expectation function, and it measures the variable's importance in affecting a continuous or a binary response. We extend previously published results to show that the defined GVIM can be represented as a function of the Conditional Average Treatment Effect (CATE) for any kind of a predictor, which gives it a causal interpretation and further justification as an alternative to classical measures of significance that are only available in simple parametric models. Extensive set of simulations using realistically complex relationships between covariates and outcomes and number of regression techniques of varying degree of complexity show the performance of our proposed estimator of the GVIM.
 
\end{abstract}


\tableofcontents
\listoftables
\listoffigures

\newpage


\section{Background}
The need for interpretability in machine learning models is an inevitable consequence of their increasing use in many scientific fields for the last few decades. Unlike in parametric statistical models, deciphering the importance of a predictor in explaining an outcome from machine learning models can be quite challenging. Often, in scientific research, parametric statistical models are preferred over black box models since they are easily interpretable. Although some work has been published\citep{molnar2020interpretable, breiman2001random} to interpret relationship between a predictor and an outcome from machine learning methods, they are mostly algorithmic and do not have any statistical or causal interpretation, which limits their use in many scientific domains, e.g., in clinical and public health. In this study, we propose a novel approach to identifying the importance of predictors for a continuous outcome from any machine learning model.

 \subsection{Related Work}
 In the last few decades, some developments have been made to explain outputs from black box machine learning models\citep{molnar2020interpretable}. The popular methods include the Variable Importance Metric (VIM)\citep{breiman2001random}, LIME (Local Interpretable Model-agnostic Explanations)\citep{ribeiro2016should}, SHAP (SHapley Additive exPlanations)\citep{lundberg2017unified} and permutation-based approaches. For example, in a recent study\citep{weller2021predicting} the authors were able to find risk factors for suicidal thoughts and behavior using multiple machine learning models using SHAP. However, it is still not clear if these methods can quantify a causal relationship between the predictors and outcomes. Breiman (2001)\citep{breiman2001random} proposed the VIM for random forests to identify important variables for prediction. The VIM provides the ranking of the predictors based on changes in prediction error. The VIM approach is widely applicable and can be effectively used for conventional linear models, generalized linear models, and generalized additive models\citep{gregorutti2017correlation, fisher2019all, hooker2021unrestricted}. Numerous studies however, have found that the permutation-based importance of random forests can produce diagnostics that are highly misleading, particularly when there is strong dependence among features\citep{strobl2007bias, strobl2008conditional, hooker2021unrestricted}. Using simulation studies \citet{hooker2021unrestricted} advocated the use of alternative measures such as conditional permutation based VIM for interpreting black box functions, instead. \citet{kaneko2022cross} has proposed a cross validation based technique to calculate VIM that minimizes the effect of feature correlations using simulations. However, most of these simulation scenarios were built on assuming the functional relationship between outcomes and predictors is linear and/or additive. It is unclear, though, how VIM would perform in more complex scenarios, such as non-linear and non additive functional relationships between predictors and outcome. Furthermore, the question still remains about the use of such a variable importance measure and whether it is possible to use this measure to discern the causal relationship (if any) between the predictor and the outcome or if the VIM can be represented as a statistical parameter. \citet{diaz2015variable} argued that variable importance and prediction have different goals, and the VIM proposed by Breiman does not have any clinical relevance. \citet{van2011targeted} created a new definition of variable importance based on a targeted causal parameter of interest, which has been very effectively used for many real life applications. \citet{fisher2019all}, defined Model Reliance (MR) based on the idea of VIM. \citet{gregorutti2017correlation, hooker2021unrestricted} have also shown similar representations of such statistical parameter. \citet{fisher2019all} further showed such a method can be represented as a statistical parameter and also have a causal interpretation. They further proposed techniques for estimating MR and their probabilistic bounds. MR was based on some pre-defined models or a model class. It is unclear if this MR or VIM can be represented as a statistical parameter in cases where the true relationship between an outcome and the predictors is completely unknown. \\
 
 In this study, we defined a new generalized variable importance metric (GVIM) at the population level. This method is a generalization of the original VIM defined by \citet{breiman2001random}.  Specifically, we
 \begin{enumerate}[(i)]
 	\item explicitly defined a population-level model-agnostic variable importance metric by generalizing Breiman's\citep{breiman2001random} VIM and Fisher's\citep{fisher2019all} MR. This generalized version of VIM (GVIM) can also be represented as a causal parameter. We focused on defining the GVIM based on the true conditional expectation of a continuous outcome given the predictors, which is not known in real life. That is, the definition is not dependent on any pre-defined functional relationship between the outcome and the predictors.
 	\item developed an estimation technique for the defined GVIM. The estimators of GVIMs are calculated using split-sample techniques as defined by \citet{buhlmann2002analyzing}. A machine learning model is first fitted with a training set, and then the GVIM is estimated using the prediction error from an independent validation set.
 	\item evaluated the statistical properties of of GVIM.
 \end{enumerate}
 
 We showed the statistical properties of GVIM based on simulations. All of the previous studies demonstrated their findings based on nonlinear and non additive models, which can be considered simpler scenarios. In the simulations of this study, we focused on scenarios that are complex, non-linear, and non-additive as detailed in Section \ref{sec.Sim}.

\section{Methods}

  \subsection{Notions of Importance}

Although there have been many algorithms established to identify variable importance, before defining any new approach, it is crucial to understand the notions of the term ``importance''. The term ``importance'' can be ambiguous without any context. Thus at first it is essential to understand the notions of importance. In general, importance of a predictor can be represented with three different notions as described by \citet{jiang2002quasi} and \citet{zhao2019causal}:

\begin{enumerate}[(i)]
	\item The first notion is to use a parametric model. Let $Y = \beta_{0}+\sum_{j=1}^{P}\beta_{j}X_{j} + \epsilon$. Here, the parameter $\beta_{j}$ can be considered as importance of $X_{j}$. The larger the the value of $\beta_j$ the more important $X_j$ is for the outcome $Y$. The p-values related to the parameters can be used as measure of importance, which is used to define the statistically significance.
	\item The second notion is to calculate the importance of a predictor $X_{j}$ by its contribution to predictive accuracy. Like the VIM for random forest (\citet{breiman2001random}) this type of notion is used when the prediction models cannot be represented by statistical parameters such as $\beta$. Inherently, the models have a non-linear form which can be represented as,
	\begin{equation*}
		\mathbb{E}(Y|X) = f(X_{1}, X_{2}, ..., X_{P})
	\end{equation*}
	such as random forests, boosting and ANNs.
	\item The third notion is causality. This is calculated by observing a change in outcome $Y$ by applying a change in intervention $X_{j}$ (change the value of $X_{j}$ from $a$ to $b$ keeping the other variables fixed). This notion is developed by \citet{van2011targeted,zhao2019causal}.
\end{enumerate}

The first notion is widely understood and used by both the statistical and broader scientific communities. The importance of predictors is obtained from parametric statistical models in the form of statistical significance. These parametric methods do not perform well if the underlying model assumptions are violated. The second notion is widely accepted in machine learning literature, is more flexible towards non-linear and/or non-additive models, and does not require any parametric assumptions. The limitations of such techniques are that they cannot always be interpreted causally\citep{van2018targeted,diaz2015variable}. The third notion of causality has received popularity among biostatisticians and clinicians in the last few decades\citep{gruber2010application}, but those measures either depend on parametric assumptions (marginal structural models) or intensive computation (TMLE)\citep{van2011targeted}. Estimation techniques and inferential tools are well-established for parametric statistical models and causal models. Estimation techniques or inference mechanisms for VIM-like methods have only been studied recently by \citet{fisher2019all}. The other popular methods, such as LIME and SHAP, although can be defined but is difficult to interpret at causally. One important finding from \citet{fisher2019all} was that the model reliance (MR), as defined by the authors, can be rewritten as a quadratic function of the expected conditional average treatment effect (CATE) for binary predictors and treatments. The model reliance defined by the authors was a generalization of VIM. Both VIM and MR are defined by applying the permutation technique to the prediction errors. \\

\subsection{Generalized Variable Importance Metric at the Population Level}

The variable importance metric (VIM) developed by \citet{breiman2001random} uses a permutations to investigate the importance of a predictor. However, this metric does not have a population-based definition and was specifically developed for Random Forests. Since, then many authors have defined the metric at the population level\citep{fisher2019all, gregorutti2017correlation, rinaldo2019bootstrapping}.  \citet{fisher2019all} redefined the VIM as ``Model Reliance (MR)''. This method extends the idea of Breiman's VIM to the population level. The authors defined MR based on a predefined prediction function $f \in \mathcal{F}$, where $\mathcal{F}$ is a predefined class of prediction functions. Our approach is very similar to that of \citet{fisher2019all}, with the important difference being that we define the metric based on the true conditional expectation function $f_0$, which is completely unknown and thus, may not belong to the class $\mathcal{F}$, i.e., $f_0 \notin \mathcal{F}$. The definition of the Generalized Variable Importance Metric (GVIM) is as follows:

\begin{enumerate} 
	\item Let outcome be $Y$ and a predictor vector $\mathbf{X} = (X_1, X_2, ..., X_P)\in \mathbb{R}^P$
	\item Let $f_0(\mathbf{X}) = \mathbb{E}(Y\mid \mathbf{X})$ be the true conditional expectation. 
	\item The GVIM can be defined as follows:
	\begin{equation}
		GVIM_{j} =\mathbb{E}_{X_j}\mathbb{E}_{\mathbf{X_{-j}}}\mathbb{E}_{ Y\mid \mathbf{X}}\left[ \left(Y - f_{0}\left(\mathbf{X}^{\prime}_{(j)}\right)\right)^{2} \right] -  \mathbb{E}_{\mathbf{X}}\mathbb{E}_{Y|\mathbf{X}}\left[\left(Y - f_{0}\left(\mathbf{X}\right)\right)^{2} \right]
		\label{pop.vim}
	\end{equation}
	Here, $\mathbf{X}_{(j)^\prime} = (X_{1}, X_{2}, ..., X_{j^{\prime}}, ..., X_{P})$ where $X_{j^{\prime}}$ is an independent replicate of $X_j$ from the marginal distribution distribution of $X_{j}$. Here $f_0$ is still the true conditional expectation function and $\mathbf{X}_{-j}$ is predictor vector without the predictor $X_j$.
	\item A variable $X_j$ can be considered as important for predicting $Y$ if by breaking the link between $X_j$ and $Y$ (that is replacing the values of $X_{j}$ with an independent replicate) the prediction error increases. The marginal distribution of $X_{j}$ remains intact.
\end{enumerate} 

Our metric is distinct from the previously developed VIMs since, it is defined using the true conditional expectation function $f_0$ for which the functional form is unknown.

\subsection{Expressing GVIM as a Causal Parameter for multinomial and continuous outcomes}

\citet{fisher2019all} showed that the MR defined using the true conditional expectation $f_0$ can be represented as a function of CATE squared for a binary treatment. In this section we show that the same relationship exists for a multinomial and a continuous predictor. To establish the relationship between GVIM and CATE let $O = (Y, X, Z)$ be random variables with outcome $Y$, exposure $X$ and confounder $Z$. \citet{fisher2019all} termed their metric as model reliance (MR), since the importance is learned from a pre-defined model. They later defined a model class reliance (MCR), which is obtained from MRs of different models and thus provides a probabilistic bound for the importance of a predictor. \citet{fisher2019all} further showed a connection between MR to conditional average treatment effect for a binary treatment. They assumed a binary treatment indicator and showed that the MR can be expressed as a function of treatment variance and conditional average treatment effect. The advantage of such a measure is that even if the treatment effect varies by sub-populations, the conditional average treatment effect may reduce to zero, but the MR can still identify the importance of the treatment. In this study we used a similar approach in defining the GVIM. Let's assume that the prediction function is given by $f_0(X, Z) = \mathbb{E}(Y\mid X, Z)$, which also the true conditional expectation. Two independent observations from $P_{O}(.)$  are $O^{(a)} = \{Y^{(a)}, X^{(a)}, Z^{(a)}\}$ and $O^{(b)} = \{Y^{(b)}, X^{(b)}, Z^{(b)}\}$. The expected squared error loss for model $f_0$ using $O^{(a)}$ is defined as

\begin{equation}
	e_{\text{orig}}(f_0) = \mathbb{E}_{X,Z,Y}\left(\left(Y^{(a)} - f_0( X^{(a)}, Z^{(a)})\right)^{2}\right)
	\label{orig}
\end{equation}
%

The VIM procedure developed by Breiman was based on permutations of the variable of interest (treatment) from the observed data. However, permutation cannot be defined at the population level. The concept of permutation can be mimicked by switching two independent replicates of the defined random vector $\mathcal{O}$ as motivated by \citet{fisher2019all}. We have assumed that $O^{(a)} = \{Y^{(a)}, X^{(a)}, Z^{(a)}\}$ and $O^{(b)} = \{Y^{(b)}, X^{(b)}, Z^{(b)}\}$ are two independent replicates from the population. To calculate the importance of the treatment $X$,  $X^{(a)}$ is switched with $X^{(b)}$ in $O^{(a)}$ and the squared loss is recalculated. The switched loss function can be written as,

\begin{equation}
	e_{\text{switch}}(f_0) = \mathbb{E}_{X^{(b)},X^{(a)}, Z,Y}\left(\left(Y^{(a)} - f_0( X^{(b)}, Z^{(a)})\right)^{2}\right)
	\label{switch}
\end{equation}

Then the generalized variable importance metric (GVIM) can be defined as

\begin{equation}
	GVIM_{X}(f_0) = e_{\text{switch}}(f_0) - e_{\text{orig}}(f_0) 
	\label{vim}
\end{equation}

We define $Y_{i}$ and $Y_{j}$ to be the potential outcomes under treatments $X = i$ and $X= j$. The conditional average treatment effect can be expressed as, $\text{CATE}_{ij}(z)^{2} = \mathbb{E}\left(Y_{i} - Y_{j}\mid Z = z\right)^{2}$. We assume
the strong ignorability of the treatment assignment mechanism, which states that, $0 < P(X\mid Z = z) < 1$ (positivity) and $(Y_{i}, Y_{j} )\perp X\mid Z$ (conditional ignorability), for all values of $Z=z$. \citet{fisher2019machine} showed that when $X$ is a binary treatment variable, then the variable importance can be defined as,  

\begin{equation}
	GVIM_{X}(f_0) =  \text{Var}(X)\sum_{x\in\{0,1\}}\mathbb{E}_{Z\mid X=x}(\mathbb{E}(Y_1 \mid Z) - \mathbb{E}(Y_0\mid Z))^2
\end{equation} 

\noindent where, $Y_1$ and $Y_0$ are potential outcomes respectively for $X = 1$ and $X = 0$. Here, $\mathbb{E}(Y_1\mid Z) - \mathbb{E}(Y_0\mid Z)$ is the conditional average treatment effect (CATE). Thus, it was shown that the GVIM based on switching is related to the CATE which is a causal parameter. In this section the aim is to investigate the relationship between GVIM and CATE for multinomial and continuous treatments. The proofs of Theorem 1 and 2 are provided in the Appendix section. \\

\begin{theorem}
	Let's the treatment be multinomial with $K$ categories ($X \in \{1, 2, ..., K\}$). That is Now with respect to the true conditional expectation $f_{0}$  then \eqref{vim} can be re-written as,
	
	\begin{equation}
		\begin{split}
			GVIM_{X}(f_0) = & e_{\text{switch}}(f_0)  - e_{\text{orig}}(f_0) \\
			= &  \sum_{k\neq j}p_{k}p_{j}\mathbb{E}_{Z\mid X=k}(\mathbb{E}(Y_k \mid Z) - \mathbb{E}(Y_j\mid Z))^2
			\label{vim.1}
		\end{split}
	\end{equation}
\end{theorem}

This shows that the GVIM can be expressed as a product of marginal probabilities of treatment levels $k$ and $j$. Under the positivity assumption, if $GVIM_{X}(f_0) = 0$, then $\mathbb{E}_{Z|X=k}\text{CATE}_{kj}(z)^{2} = 0$, for all $k$. Thus the treatment is not important for outcome $Y$. If $GVIM_{X}(f_0) > 0$, then the treatment is effective. In the case of heterogeneous treatment effect, this measure can still recognize an important treatment variable\citep{fisher2019all}. The GVIM has causal interpretation with respect to the true conditional expectation $f_0$ under the positivity and consistency assumptions. In the next step the focus is to investigate  the decomposition when the treatment of interest is continuous.  \\

\begin{theorem}
	For a continuous treatment \eqref{vim} can be rewritten as,
	\begin{equation}
		GVIM_{X}(f_0)   = \mathbb{E}_{X^{(b)}}\mathbb{E}_{X^{(a)}}\mathbb{E}_{Z|X^{(a)}} \left(\mathbb{E}\left(Y_{X^{(a)}} - Y_{X^{(b)}} | Z\right)\right)^2
		\label{VIM.cont}
	\end{equation}
\end{theorem}

that for a continuous treatment the GVIM can be expressed as a function of conditional average treatment effect.

\subsection{Estimating GVIM}
\label{subsec.Est}

\subsubsection{Motivation}
\label{subsubsec.Est.mot}
GVIM is not just a statistical parameter but, under certain assumptions, can be represented as a causal parameter. The GVIM is defined at the population level using the true conditional expectation $f_0$, which then can be expressed as a function of conditional average treatment effect averaged over all possible treatments. Since the defined quantity depends on the true conditional expectation, which has an unknown functional form, our approach is to build a model agnostic estimator based on black box machine learning models.  Recall the functional form of the GVIM defined in \eqref{pop.vim}. The switched error can be decomposed as following,
\begin{equation}
	\begin{split}
		e_{\text{switch}}(f_0)=	& \mathbb{E}_{X_j}\mathbb{E}_{\mathbf{X_{-j}}}\mathbb{E}_{ Y\mid \mathbf{X}}\left[ \left(Y - f_{0}\left(\mathbf{X}^{\prime}_{(j)}\right)\right)^{2} \right]  \\
		=	& \mathbb{E}_{X_j}\mathbb{E}_{\mathbf{X_{-j}}}\mathbb{E}_{ Y\mid \mathbf{X}}\left[ \left(Y - f_0(\mathbf{X}) + f_0(\mathbf{X}) - f_{0}\left(\mathbf{X}^{\prime}_{(j)}\right)\right)^{2} \right]  \\
		= &   \mathbb{E}_{\mathbf{X},Y}\left[\left(Y - f_{0}\left(\mathbf{X}\right)\right)^{2} \right] +  \mathbb{E}_{X_j}\mathbb{E}_{\mathbf{X_{-j}}}\left[ \left( f_0(\mathbf{X}) - f_{0}\left(\mathbf{X}^{\prime}_{(j)}\right)\right)^{2} \right] 
	\end{split}
	\label{pop.vim.decompose}
\end{equation}

Here , $Y = f_{0}(\mathbf{X}) + \varepsilon$ and $\varepsilon \perp \mathbf{X}$ is some random error. Then, as defined in \eqref{orig} and in \eqref{pop.vim.decompose}, $ e_{\text{orig}}(f_0) =   \mathbb{E}_{\mathbf{X},Y}\left[\left(Y - f_{0}\left(\mathbf{X}\right)\right)^{2} \right]  = \text{Var}(\varepsilon) = \sigma_{\varepsilon}^2 = e_{\text{orig}}$, is the variance of the error term $\varepsilon$, which is the irreducible error. Furthermore, the term, $\mathbb{E}_{X_j}\mathbb{E}_{\mathbf{X_{-j}}}\left[ \left( f_0(\mathbf{X}) - f_{0}\left(\mathbf{X}^{\prime}_{(j)}\right)\right)^{2} \right] $ represents the GVIM. This term indicates how much the prediction value changes if we replace the variable $X_{j}$ with it's independent replicate $X_{j^{\prime}}$, without changing any other variables. \citet{gregorutti2015grouped} have shown that for additive functions \eqref{pop.vim.decompose} can be simplified as $2\text{Var}_{X_j}(f_j(x_j))$, where $f_j(.)$ is the relationship between $X_j$ and $Y$. In case of linear functions this further simplifies to $2\beta_j^2\text{Var}_{X_j}(X_j)$, which is exactly twice of the \emph{leave one covariate out (LOCO)} estimator of variable importance\citep{rinaldo2019bootstrapping, verdinelli2023feature}.  \\

To create a model agnostic estimation technique for the GVIM, we followed a similar approach as was defined by \citet{breiman2001random} and \citet{fisher2019all}. We assume that we have a finite dataset with $n$ independent observations and $p$ predictors. The causal parameters are generally estimated using all the observations in the sample dataset. This method is appropriate for parametric models or TMLE given that the sample size is sufficiently large. However, it has been well established that the in-sample prediction errors or prediction errors based on training sets are an underestimation of the true error\citep{efron1997improvements,tibshirani1993introduction}. \citet{rinaldo2019bootstrapping} discussed the trade-off between prediction accuracy and inference. It was elucidated that splitting increases the accuracy and robustness of inference but can decrease the accuracy of the predictions. While Bagging\citep{breiman1996bagging} is effective for Random Forests, however, can produce unstable predictions when applied to other machine learning models\citep{buhlmann2002analyzing}. Thus, we decided to use subbagging (subsample aggregation) as described by \citet{buhlmann2002analyzing} . Furthermore it is crucial to know the behavior of such metric when the model is mis-specified. For example the relationship with predictors and the outcome could be non-linear and non additive, but to better explain the model a researcher may end up fitting an additive and/or linear model. In such case of model mis-specification the permutation based importance metric may not show the true importance of a predictor and provide a biased estimate of the GVIM. All the studies conducted so far have focused on fitting the true model or a predefined model. Our aim in this study is to investigate on various additive and non-additive models to evaluate how the model mis-specification can affect the estimation of GVIM.

\subsection{Estimating GVIM: Training versus validation set}
\label{short.sim}

In majority of the studies\citep{fisher2019all, hooker2021unrestricted, strobl2007bias, strobl2008conditional} the permutation based importance measure was estimated utilizing training error. This can result in over or underestimation of the true GVIM when the model is overfitted. In this section we provide a simple scenario where we showed that the GVIM calculated using the training error can provide a over or underestimation of the true GVIM. We generated the data using the following model:

\begin{equation}
	Y =  X _1+ 2X_1^2 + 2X_2 + \mathbf{0^{\prime}Z}+ \epsilon
\end{equation}

Here $\epsilon \sim \text{N}(\mu = 0, \sigma^2 = 25)$ and $\epsilon \perp X_1, X_2$.  That is the conditional expectation $\mathbb{E}(Y|X_1, X_2)$ is expressed as a quadratic function of $X_1$ and a linear function of $X_2$. Here, the irreducible error was chosen to be very large ($e_{\text{orig}} = \sigma^2 = 25$). Both the predictor variables $X_1, X_2$  were generated from independent standard normal distribution. Another random vector $\mathbf{Z}$ was generated from a normal distribution which did not have any relationship with $Y$. Here $\mathbf{Z}$ is a predictor matrix were the number of variables were varied between 1 to 40. The length of this vector of nuisance predictors increases the complexity of the training model.  Since, the target was to investigate the GVIM estimates in the existence of overfitting, we only generated 100 samples. In this simulation we calculated the GVIM by fitting the following models,
\begin{align}
	E(Y\mid X_1 X_2, \mathbf{Z}) & = \beta_0 + \beta_1X_1 + \beta_2X_2 + \boldsymbol{\beta}_z^{\prime}\mathbf{Z} 		\label{over.fit1} \\
	E(Y\mid X_1 X_2, \mathbf{Z}) & = \beta_0 + \beta_1X_1 + \beta_{12}X_1^2 + \beta_2X_2 + \boldsymbol{\beta}_z^{\prime}\mathbf{Z} 
	\label{over.fit2}
\end{align}

\begin{figure}[H]
	\centering
	\includegraphics[scale = 0.75]{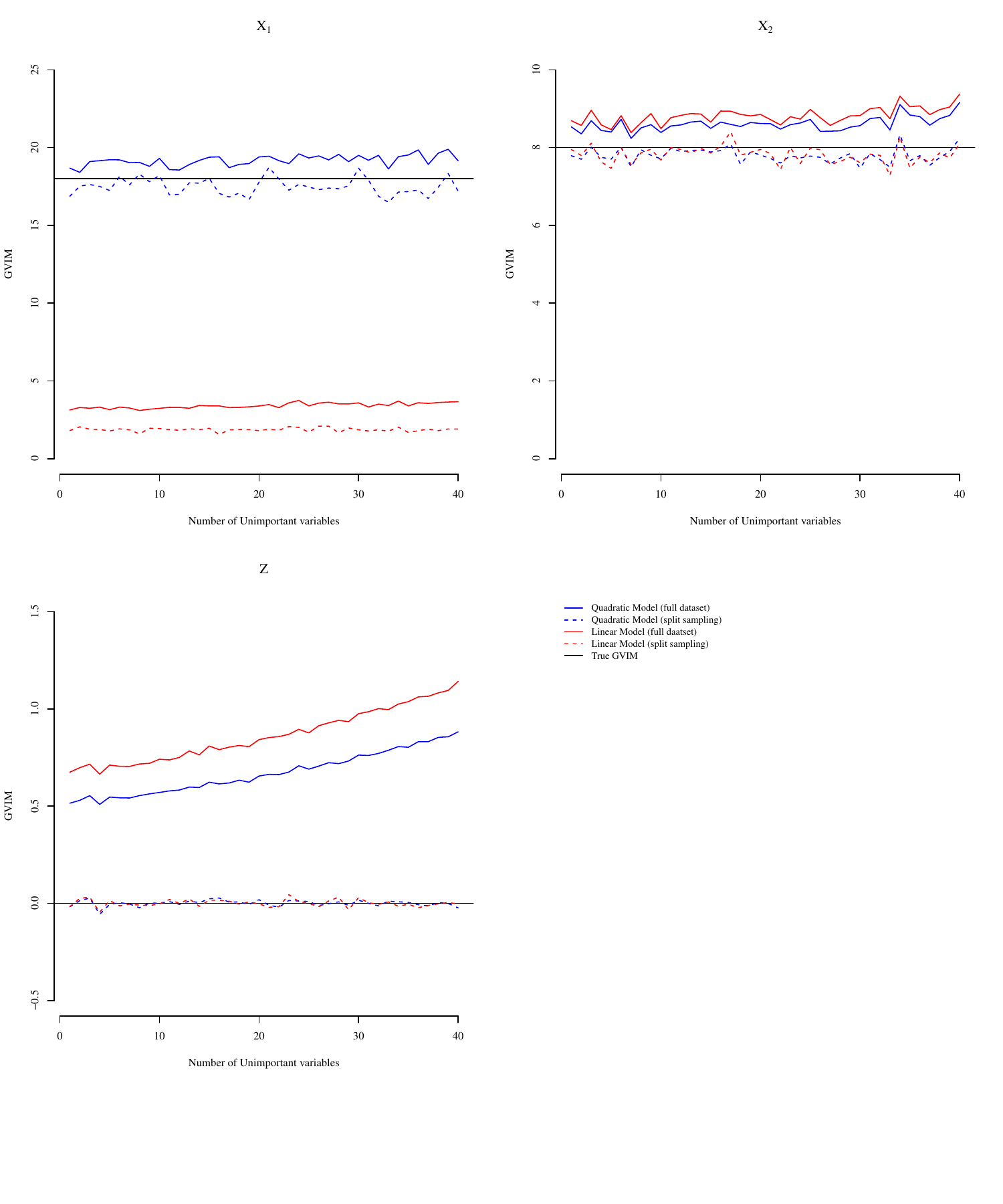}
	\caption{VIM vs number of unimportant variables for a small training and validation set}
	\label{FigA1}
\end{figure}

Here the equation \eqref{over.fit2} fits the model with the correct functional form of both $X_1$ and $X_2$. We generated 500 datasets for each size of $\mathbf{Z}$. The GVIMs of the predictors were calculated in two different approaches. In the first approach the full dataset was used to fit the model and to estimate the GVIMs, i.e., without using split sampling. In the second approach the GVIMs were calculated using the split sampling technique. In this approach two thirds (66\%) of the data was used as the training set and remaining one third of the data was used for the validation set. The GVIMs were then calculated from the validation set error after fitting the model with the training set.  Then the average GVIM estimates from both the approaches were calculated from the 500 simulations. Figure \ref{FigA1} show the behavior of the GVIMs as we change the number of unimportant predictors. As can be seen from the Figure \ref{FigA1}, that for the variable $X_1$ the GVIM estimated from the full set for the quadratic model (\eqref{over.fit2}) overestimated the true GVIM.  The linear model (\eqref{over.fit1}) was not able to identify the importance of $X_1$, since the model was mis-specified.  Similarly for the variable $X_2$, the GVIMs calculated using split sampling from both the quadratic and the linear model had smaller absolute distance from the true GVIM. The full datasets for both the models were overestimating the GVIM by significant margins as the number of nuisance covariates increased. GVIM was also estimated for all the nuisance predictors and then the average of the GVIM estimates was calculated. As can be seen from the Figure \ref{FigA1}, when GVIM was estimated from the full datasets, then both the linear and the quadratic models were overestimating the GVIMs. Also, the GVIM was increasing with the number of unimportant nuisance when calculated using the full datasets. However, when the GVIMs were estimated using the validation error then the estimates remained very close to 0. These results confirms that when a model is overfitted, estimating GVIM using the training error produces biased results, specially, for the nuisance predictors.

\section{Investigating the statistical properties of GVIM}
\label{sec.Sim}
We assessed the properties of GVIM estimator using simulations which allows us to compare the estimators to the true GVIM values in the population. The simulations were constructed to reflect the complex scenarios encountered in various scientific fields such as, clinical and public health. For this simulation study we constructed the response $Y$ using the following model inspired from \citet{friedman1991multivariate},
\begin{equation}
		\begin{split}
		Y  = & 2X_1 - 4X_1C_1 + 2C_1 + 2\log(\mid X_2 X_3\mid)  + (X_4 - 0.5)^3 - 2X_5 + 2\sin(\pi U_1U_2)  \\
		& - \mathcal{I}(C_2 = 2) + 2 \mathcal{I}(C_2 = 3)    + \epsilon
	\end{split}
	\label{sim.mars}
\end{equation}

The coefficients were selected arbitrarily. Here, $\epsilon \sim \text{N}(0,1)$ is the random error. The continuous random predictors $X_2, X_3, X_4, X_5$ were generated from $\text{Normal}(0,1)$ distributions. The categorical predictors  $C_1$ and $C_2$ were generated from multinomial and binomial distributions respectively, which were also independent. The variables  $U_1$ and $U_2$ were drawn from Uniform(-1,1) distribution, which were independent of the other predictors. Another 45 variables $X_8 - X_{52}$, were generated from $N(0,1)$, These predictors were not related to the outcome, or the other predictors, and thus had no importance. The predictor $X_1$ was generated using the following model,

	\begin{equation}
	X_1  =  -0.5 + C_1 -0.5X_2 + 0.5X_3 + 0.3X_4 - 0.3X_5  + \nu; \text{ }\nu \sim N(0,\sigma^2_x = 0.066)
	\label{x.gen}
\end{equation}

this equation was used to make sure that the expectation, $\mathbb{E}(X_1) = 0$, and the marginal variance $\text{Var}(X_1) = 1$. Here, we are interested in identifying the importance of every variable. The purpose of using such an equation \eqref{sim.mars} was to make sure that we have predictors with linear, polynomial, and oscillating functional relationships with the outcome $Y$. We further ensured that there was interaction between $X_1$ and $C_1$.  We generated training sets of sizes 50, 100, 200, 500, 1000, 5000, 10,000 and 50,000. The aim of this simulation was to investigate how accurately a specific method can estimate GVIM of the predictors $X_1-X_5, C_1, C_2, U_1, U_2$. At first to calculate the true GVIM estimates, we generated all the predictors of size $n_{\text{pop}} = 100,000$. This large dataset was considered as an empirical population, from which the true values of the GVIMs were calculated using Monte Carlo expectations. The true $e_{\text{orig}} = \text{Var}(\epsilon) = 1$ for the dataset and the $e_{\text{switch}}$s varied by the predictors. The true GVIMs for each important predictor are reported in the Table \ref{tab.1}.  In the next step we generate training sets of various sizes and fitted the following models, 
\begin{enumerate}[(a)]
	\item The \emph{oracle model}, where the model was fitted with the correct functional forms based on the conditional expectation of the predictors using the seven important predictors. The continuous  predictors were then transformed to the functional forms defined in equation \eqref{sim.mars}. That is we first performed the following transformations: $Z_1 = \log\left(\mid X_2 X_3 \mid\right), Z_2 = (X_4 - 0.5)^3$ and $Z_3 = \sin(\pi U_1 U_2)$ and then fitted a linear model on $Y$ using the following model. 
	\begin{equation*}
		E(Y\mid X_1, C_1, Z_1, Z_2, X_5, C_2) = \beta_0 + \beta_1 X_1 + \beta_{12}X_1C_1 + \beta_3Z_2 + \beta_4Z_2 + \beta_5X_5 + \beta_6 Z_3 + \beta_6\mathcal{I}(C_2 = 2) + \beta_7 \mathcal{I}(C_2 = 3)
	\end{equation*}
	\item A  \emph{GAM} model was fitted with cubic splines using purely additive structure. For the categorical predictors we used the linear terms without considering any interaction with the other predictors. 
	\item A  \emph{XGBoost} model proposed by \citet{chen2016xgboost} was used as an example of a black-box model. The number of boosting iterations was chosen to be between 100 trees to 5000 trees, the maximum depth of a tree was set to be between 2-6, and the learning rate was varied between 0.05-0.3. These hyper-parameters were chosen based on results obtained during 200 simulations separately for each training size. The hyper-parameters were then varied based on the training size. For the smaller training sizes the learning rate was chosen to be very small (0.05), with large number of trees (5000). The rate was increased along with decreasing number of trees as the training set increased, since a small learning rate with a large number of additive trees did not improve the results.
\end{enumerate}

The estimated $e_{\text{orig}}(\hat{f})$ was calculated from a validation set for a specific model using,
\begin{equation}
	\hat{e}_{\text{orig}}(\hat{f}) = \dfrac{1}{n_v}\left(\sum_{i=1}^{n_v}y_{i} - \hat{f}(\mathbf{X})\right)^2
\end{equation}
where $\mathbf{X} = (X_1, X_2, ..., X_{52}, C_1, C_2, U_1, U_2)^{\prime}$ is the predictor vector which includes all the predictors (both important and non-important). The $e_{\text{switch}}$ was then estimated by,
\begin{equation}
	\hat{e}_{\text{switch}}(\hat{f}) = \dfrac{1}{n_v}\left(\sum_{i=1}^{n_v}y_{i} - \hat{f}(\mathbf{X}^{\prime}_{(j)})\right)^2
\end{equation}
here, $\mathbf{X}^{\prime}_{(j)}$ is the predictor matrix where the $j^{th}$ predictor is permuted in the validation set and $n_v$ is the size of the validation set.

We simulated a dataset using the equation \eqref{sim.mars} of varying training sizes.  We fitted the models using two thirds of the data and then estimated the $GVIM_j$ for the $jth$ predictor using the remaining one third of the data. Table \ref{tab.1} shows the relative bias of the estimated GVIMs by training set sizes along with the GVIMs. We also presented the the ratio $\frac{\hat{e}_{\text{orig}}}{e_{\text{orig}}}$s in the Table \ref{tab.1}. Figures \ref{Fig1} to \ref{Fig4} show the box plot of the GVIMs over the 200 simulations by training sizes for the predictors $X_1$, $C_2$, $X_2$, $X_4$ and $X_5$. Box-plots for the remaining predictors are presented in the appendix section.

\subsection{Results from the simulations}
The Table \ref{tab.1} contains the \%Bias of the estimated GVIMs. The \%Bias  was calculated using the following formula:
\begin{equation}
	\text{\%Bias}(\widehat{GVIM}_j) = \dfrac{\widehat{GVIM}_j - GVIM_j}{GVIM_j}
	\label{bias}
\end{equation}
\begin{table}[H]
	\centering
	\caption{Table for the \%Bias in estimated GVIM for all the important predictors }
	\label{tab.1}
	\begin{tabular}{rrrrrrrrr}
		\hline
		& & & \multicolumn{2}{c}{oracle} &  \multicolumn{2}{c}{XGBoost} &  \multicolumn{2}{c}{GAM} \\
		\hline
		& Training Size & True & $\frac{\hat{e}_{orig}}{e_{\text{orig}}}$ & \%Bias & $\frac{\hat{e}_{orig}}{e_{\text{orig}}}$ & \%Bias & $\frac{\hat{e}_{orig}}{e_{\text{orig}}}$ & \%Bias \\ 
		\hline
		$X_1$ & 50 & 8.00 & 1.27 & 2.17 & 36.81 & -100.48 & 35.48 & 89.69 \\ 
   & 500 & 8.00 & 1.02 & 0.28 & 9.97 & -88.33 & 6.19 & -55.55 \\ 
   & 5000 & 8.00 & 1.01 & 0.28 & 3.04 & -83.14 & 5.47 & -57.28 \\ 
   & 50000 & 8.00 & 1.00 & 0.01 & 1.46 & -77.36 & 5.35 & -57.48 \\ 
  $C_1$ & 50 & 9.98 & 1.27 & 2.95 & 36.81 & -97.48 & 35.48 & -37.05 \\ 
   & 500 & 9.98 & 1.02 & 1.30 & 9.97 & -67.57 & 6.19 & -79.30 \\ 
   & 5000 & 9.98 & 1.01 & -0.06 & 3.04 & -33.31 & 5.47 & -80.10 \\ 
   & 50000 & 9.98 & 1.00 & -0.08 & 1.46 & -23.60 & 5.35 & -79.99 \\ 
  $X_2$ & 50 & 9.85 & 1.27 & -5.03 & 36.81 & -90.68 & 35.48 & -13.29 \\ 
   & 500 & 9.85 & 1.02 & -2.68 & 9.97 & -30.59 & 6.19 & 6.01 \\ 
   & 5000 & 9.85 & 1.01 & -0.18 & 3.04 & -3.26 & 5.47 & 9.07 \\ 
   & 50000 & 9.85 & 1.00 & 0.10 & 1.46 & 2.82 & 5.35 & 9.51 \\ 
  $X_3$  & 50 & 9.85 & 1.27 & 1.84 & 36.81 & -90.70 & 35.48 & -11.04 \\ 
   & 500 & 9.85 & 1.02 & 0.27 & 9.97 & -28.79 & 6.19 & 7.97 \\ 
   & 5000 & 9.85 & 1.01 & -0.06 & 3.04 & -2.99 & 5.47 & 9.65 \\ 
   & 50000 & 9.85 & 1.00 & 0.00 & 1.46 & 2.75 & 5.35 & 9.53 \\ 
  $X_4$  & 50 & 48.76 & 1.27 & 0.27 & 36.81 & -54.78 & 35.48 & -33.58 \\ 
   & 500 & 48.76 & 1.02 & 2.23 & 9.97 & -18.45 & 6.19 & -3.31 \\ 
   & 5000 & 48.76 & 1.01 & -0.94 & 3.04 & -4.97 & 5.47 & -3.65 \\ 
   & 50000 & 48.76 & 1.00 & -0.37 & 1.46 & -1.08 & 5.35 & -2.73 \\ 
  $X_5$  & 50 & 8.00 & 1.27 & -1.50 & 36.81 & -58.04 & 35.48 & 9.67 \\ 
   & 500 & 8.00 & 1.02 & -1.22 & 9.97 & -13.46 & 6.19 & -1.34 \\ 
   & 5000 & 8.00 & 1.01 & 0.04 & 3.04 & 0.52 & 5.47 & 0.29 \\ 
   & 50000 & 8.00 & 1.00 & 0.01 & 1.46 & 2.81 & 5.35 & 0.16 \\ 
  $U_1$ & 50 & 3.10 & 1.27 & -0.26 & 36.81 & -102.53 & 35.48 & -104.08 \\ 
   & 500 & 3.10 & 1.02 & 0.01 & 9.97 & -92.84 & 6.19 & -100.04 \\ 
   & 5000 & 3.10 & 1.01 & -0.21 & 3.04 & -56.58 & 5.47 & -99.98 \\ 
   & 50000 & 3.10 & 1.00 & 0.12 & 1.46 & -20.30 & 5.35 & -100.00 \\ 
		\hline
	\end{tabular}    
\end{table}
As can be seen from Table \ref{tab.1}, the Oracle model provided the lowest bias for all the predictors of interest, as expected. We showed the results of the predictors that were included in the equation \eqref{x.gen}. The results for the rest of the variables are provided in Appendix Table \ref{tab.A2}, since the results are unremarkable. The bias decreases as the training size increases, and the ratio  $\frac{\hat{e}_{orig}}{e_{\text{orig}}}$  also approaches 1 with increasing training size. The GAM model performs poorly in estimating GVIM for the predictors with interaction terms and also for predictors with non-linear terms ($X_2$, $X_3$, $U_1$, and $U_2$). For $X_4$ and $X_5$, which had polynomial and linear relationships with the outcome, the \%Biases were small for the GAM model. It is important to notice that since GAM was a misspecified model, it provides a very high estimate for the $e_{\text{orig}}$ even for very large training sizes. XGBoost, on the other hand, provides an estimation of the $e_{\text{orig}}$ closer to the true value of $e_{\text{orig}}$ for increasing training sizes for most of the variables. The \%Biases decreased as the training size increased for the XGBoost model for all the predictors other than $X_5$. Clearly, XGBoost provides better estimates of the GVIM compared to the misspecified GAM. However, the biases in estimating GVIM for $X_1$ and $C_1$ are substantially large. \\

\begin{figure}[H]
	\centering
	\includegraphics[scale = 0.6]{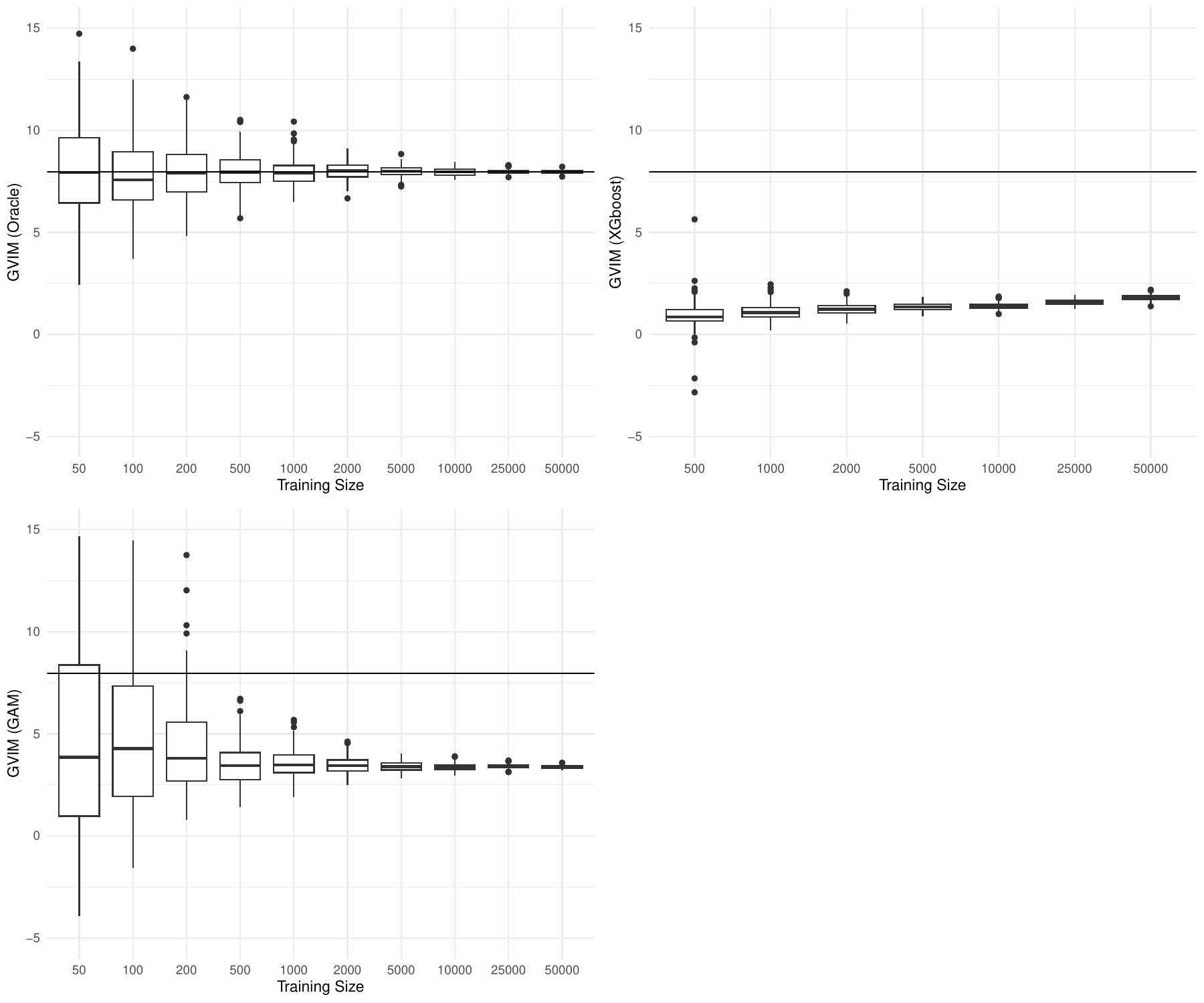}
	\caption{Estimated GVIM for the predictor $X_1$ }
	\label{Fig1}
\end{figure}

\begin{figure}[H]
	\centering
	\includegraphics[scale = 0.6]{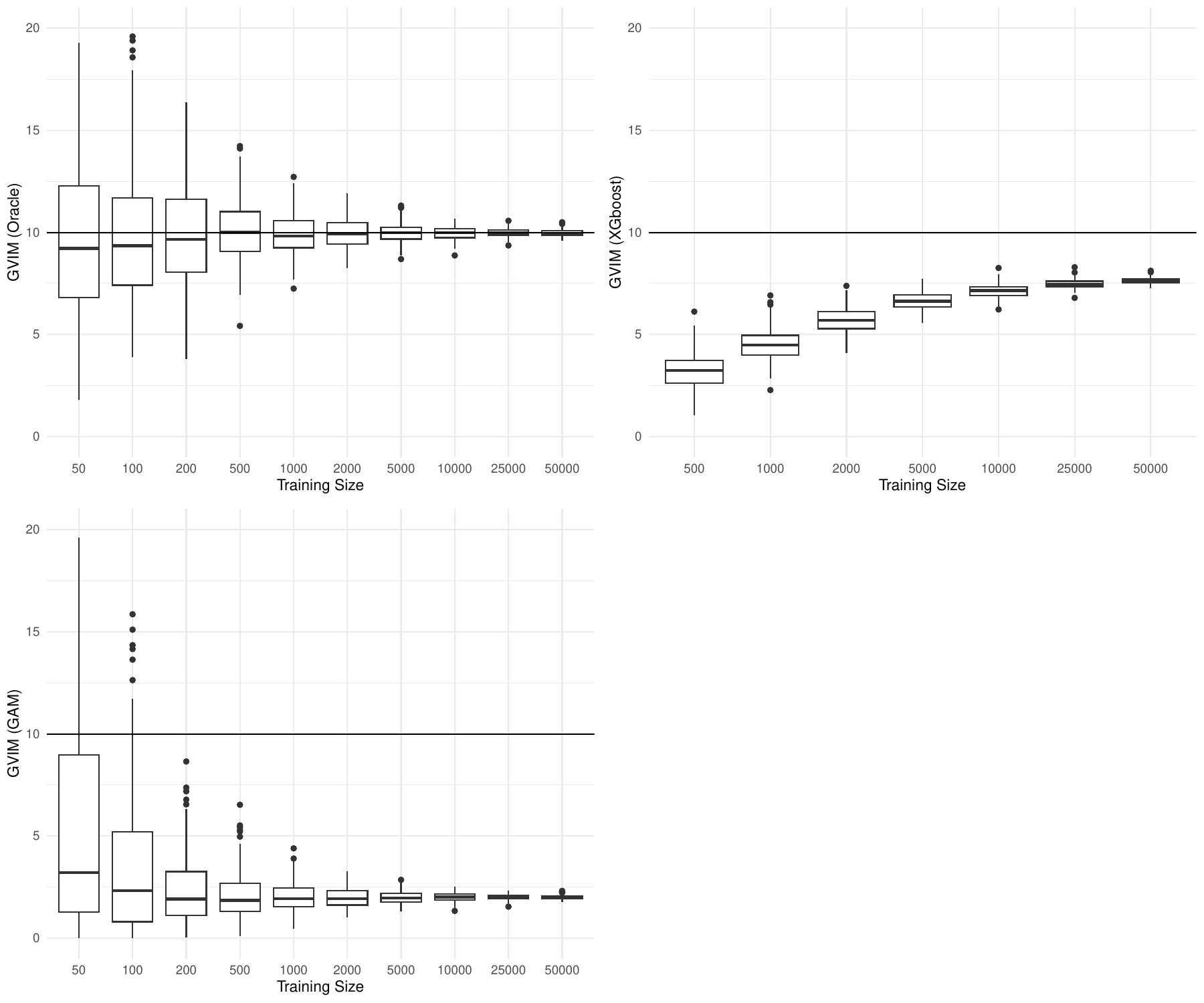}
	\caption{Estimated GVIM for the predictor $C_1$ }
	\label{Fig2}
\end{figure}

\begin{figure}[H]
	\centering
	\includegraphics[scale = 0.6]{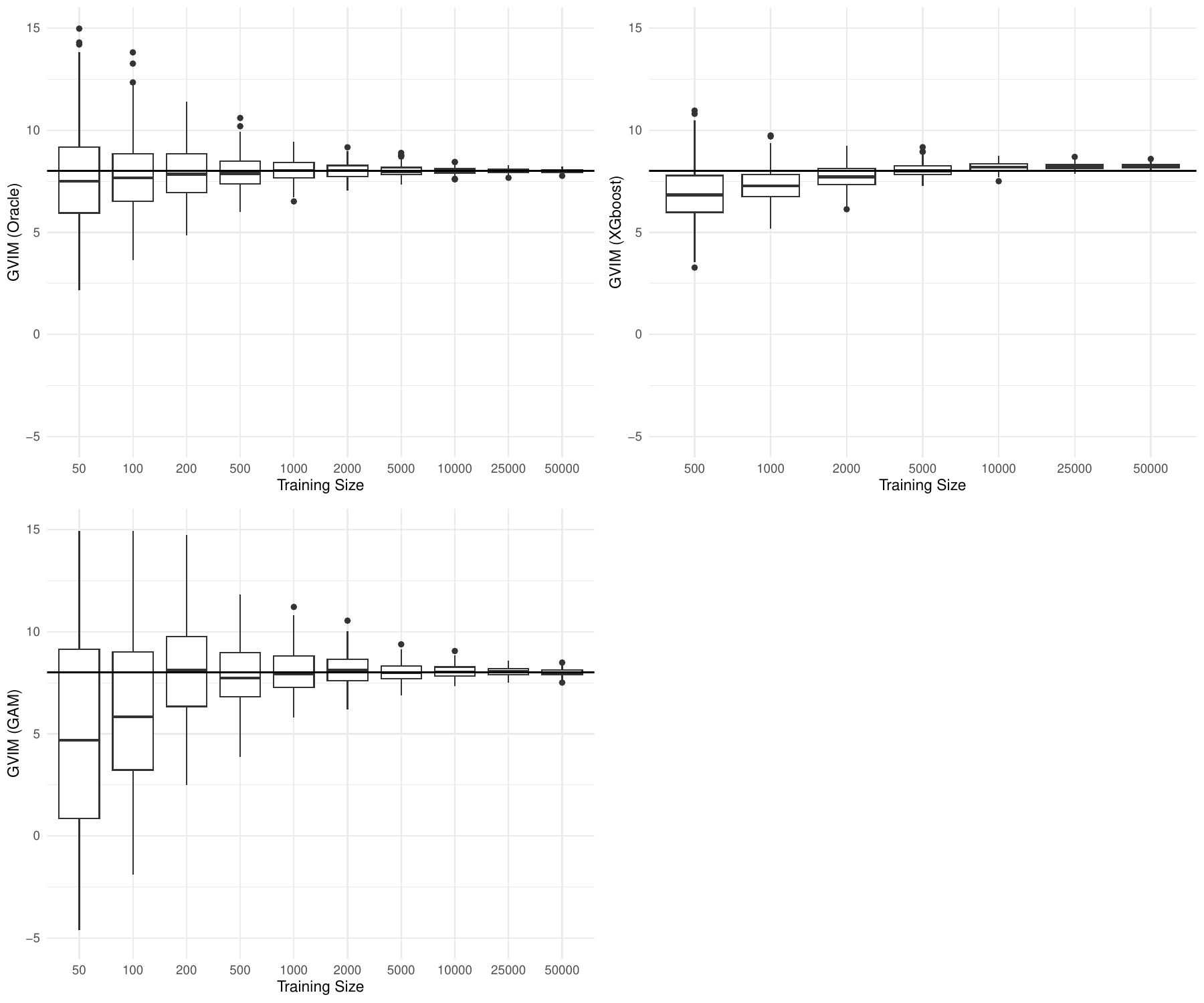}
	\caption{Estimated GVIM for the predictor $X_5$ }
	\label{Fig3}
\end{figure}

\begin{figure}[H]
	\centering
	\includegraphics[scale = 0.6]{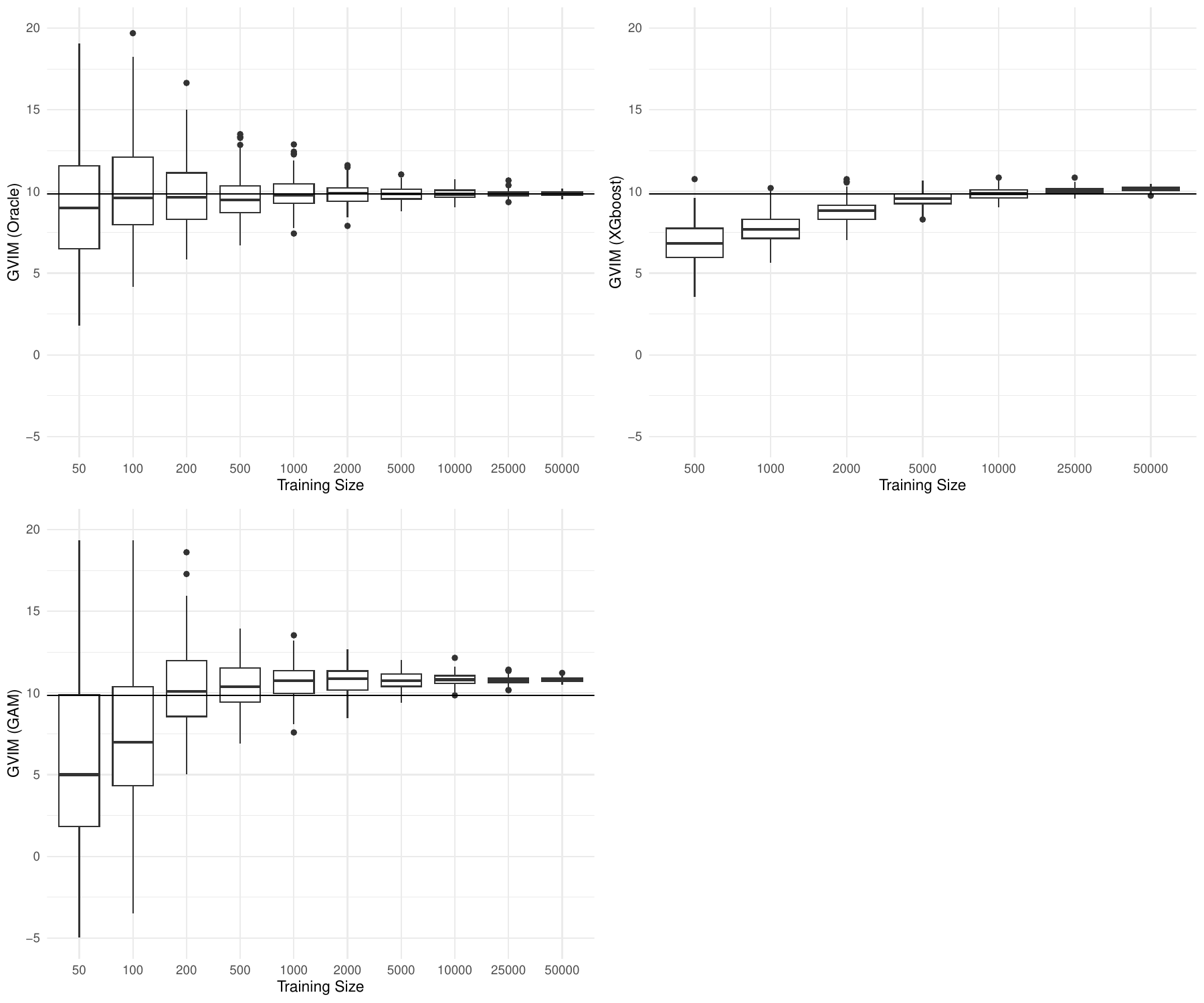}
	\caption{Estimated GVIM for the predictor $X_2$ }
	\label{Fig4}
\end{figure}

The box plots in figures \ref{Fig1} to \ref{Fig3} represent the distributions of the GVIM estimates for $X_1$, $C_1$, and $X_5$ over the 200 simulations. The plots for the rest of the variables are provided in the Appendix section. It can be observed that the Oracle model produces nearly unbiased and consistent estimators of GVIMs for all the training sizes. GAM, on the other hand, produces biased and inconsistent estimates when the functional forms are misspecified. The model produces consistent estimates of GVIM when the predictors have additive relationships with the outcome. When training size is very small (<5000) GVIMs calculated from XGBoost tend to be underestimated for all the variables. This, bias is caused by the large prediction error. For most of the predictors this bias converges to zero with increasing training size. For the predictors which had complex relationship with the outcome the bias bias was converging at a slower rate. The predictors $X_1$ and $C_1$ had interaction effect on the outcome and $X_1$ had small correlation with five other variables and as can be seen from the Figures \ref{Fig1} and \ref{Fig2}, their biases did not converge even when the training size was 50000. 

\section{Discussion}

The aim of this study was to develop a novel procedure to extract interpretability from black-box machine learning models. Our definition of GVIM is a generalization of the VIM developed by \citet{breiman2001random} and the MR developed by \citet{fisher2019all}. The most important aspect of GVIM is that its definition is model-agnostic and does not require any knowledge of the functional relationship between the predictors and the outcome. That is, unlike other methods as proposed by \citet{fisher2019all} and \citet{hooker2021unrestricted}, the model class does not have to be defined a priori. The estimation procedure is developed based on a permutation technique. One of the biggest criticisms of such procedures was that this type of estimation technique does not have any causal interpretation\citep{diaz2015variable, van2011targeted}. However, \citet{fisher2019all} has shown for binary treatments, and we further showed for multinomial and continuous treatments that once the GVIM is defined at the population level, it can then be expressed as a function causal parameter for any treatment. That is, the GVIM can extract causal interpretations using predictions from black-box models. Previous work has focused on ensemble machine learning methods such as super learner\citep{van2011targeted, van2018targeted} to estimate a pre-defined target parameter of interest. For example, the TMLE estimators\citep{van2011targeted,van2018targeted} achieve that goal in identifying the importance of a predictor/treatment. The idea proposed in this paper is to directly use the predictions from a machine learning method to evaluate the importance of a predictor without pre-defining a target parameter. The goal of this paper was to propose a properly defined metric at the population level, propose an estimator and investigate the properties of those estimators for any machine learning model. To our knowledge this is a first paper that approaches the problem of variable importance of black-box models from a classical statistical point of view: by explicitly defining a suitable population parameter with suitable properties, including a causal interpretation, and then investigating properties - like bias and consistency - of its various estimators.\\

One of the key factors to consider while estimating GVIM is whether to use the full dataset for estimation or an independent validation set. It is a well-established fact that using in-sample predictions can produce optimistic prediction error estimates. \citet{fisher2019all} argued that for large sample sizes, the split sampling technique is not required to estimate the GVIM type metric when the prediction function is assumed to be from a known class of functions. However, when the prediction functions are not known a priori, the minimum sample size that would work without split-estimation is impossible to guess, hence we recommend estimating GVIM using split sampling techniques. We presented an example in Section \ref{short.sim} where we conducted a very simple simulation to show that the GVIM can be underestimated or overestimated when estimated using the full dataset, even if a model is fitted with the true functional relationship between the outcome and the predictors if the predictor space is large and the irreducible error (i.e., the error variance) is also large. GVIM estimates from the full dataset produced biased estimates, although the split sampling technique produced unbiased estimates. Since real-life scenarios can have many important predictors along with a large number of nuisance predictors, as well as unobserved sources of variation in the outcome, it is always possible to overfit a model. The size of the validation set has little impact on the estimation procedure of the GVIM. The impact of the training set size has a larger impact on estimating the GVIM. Based on our observations from various simulation scenarios, we recommend that two thirds of the data be used for the training set and the remaining one third of the data be used for the validation set.  \\

The aim while calculating GVIM should be to fit a model that provides the prediction closest to the true conditional expectation function. Our study explored the properties of GVIM obtained from simulations which was constructed based on realistically complex relationships between the predictors and outcome. We went beyond simple scenarios as proposed by many other studies \citep{hooker2021unrestricted, strobl2007bias, strobl2008conditional} for our simulation (e.g., linear or additive prediction functions), which allowed us to investigate the effect of regression model mispecification on GVIM estimation. This can be observed from the GAM model that was used in this study. The model was fitted with all the predictors using only the spline-expanded additive terms. Since no interaction or multiplicative terms were included in the GAM prediction model, it failed to produce unbiased estimators for the GVIM when predictors had interactions or had a multiplicative affect on the outcome, while GVIMs for truly additive terms were accurately estimated. The Oracle model, on the other hand, produced nearly unbiased and consistent estimators for the GVIMs. This results are not surprising since this model was fitted using the true functional form of the predictor-response relationship. The results also show that XGBoost, our exemplar black-box model with universal approximator properties, performed much better in terms of estimating GVIM than a mis-specified GAM. Still, even XGBoost produced biased estimates for some GVIMs, especially for small sample sizes. To investigate further, we focused on the validation set size. To eliminate bias due to the small size of the validation set, we used the large dataset, which was generated to obtain near-exact estimates of original and switched prediction errors. However, the bias still persisted, indicating that the bias was mostly attributed to the training set size. For example, the bias for the predictors $X_4$, $X_5$, and $C_2$ decreased much faster to 0 compared to the other variables. Especially for $X_4$, which had a polynomial association with the outcome, the bias in GVIM from the XGBoost model decreased to 0 very sharply. The predictors $X_1$ and $X_5$, both had linear associations with the outcome and also have the same importance. The bias decreased much faster to 0 for $X_5$ compared to $X_1$, due to $X_1$ having an interaction with $C_1$ and being correlated with multiple other predictors, as can be seen in equation \eqref{x.gen}. Thus, interaction terms and correlations between the predictor variables can induce bias in estimating GVIM, especially for black-box models like XGBoost. Although the bias decreased for very large training sizes, the reductions were slower for the predictors that had an interaction or multiplicative effect on the outcome and, furthermore, had correlations between themselves. This finding from a black box model such as XGBoost is not unusual since many other studies have also reported biased estimations of permutation-based feature importance in the presence of strong dependence between the predictors. Several studies, such as \citet{hooker2021unrestricted} and \citet{strobl2007bias}, have found that the VIM was overestimated by Random Forest; however, their simulation relied on linear prediction functions with different levels of correlations between two or more predictors. \citet{verdinelli2023feature} has studied LOCO estimators and showed that the variable importance based on LOCO from random forests can have downward bias. In our simulations, we found both downward and upward bias while estimating GVIM using XGBoost. As mentioned by \citet{hooker2021unrestricted} and \citet{verdinelli2023feature}, this can be due to a black box model's inability to extrapolate in low-density regions of the joint distribution of the predictors when there is strong dependence among the predictors. \citet{hooker2021unrestricted} and \citet{fisher2019all} have proposed some alternative solutions based on conditional permutations. However, these solutions may not perform appropriately when the relationships between the outcome and the predictors are complex. Further research is needed to understand the bias-variance decomposition of the metric and to find an efficient procedure to calculate standard errors. Furthermore, this metric is defined based on squared error loss, which may not be appropriate for non-binary, non-continuous responses. This method of estimating GVIM performs poorly in the presence of a high correlation between two or more predictors\citep{gregorutti2017correlation}. In our current work, we focus on the bias-variance decomposition of the GVIM to investigate novel methods for bias correction. \\

In this study, we proposed a model-agnostic generalized version of the VIM that has a causal interpretation. We further proposed an algorithm to estimate the metric. Since this metric is defined at the population level using expectations and then estimated from the data, standard errors and confidence intervals can be defined. This approach to using predictions to draw causal inferences is different from the way machine learning methods are used in causal inference literature. Our future work will focus on redefining the metric for categorical responses and modifying the metric when predictors are highly correlated with each other. We are also working on understanding the bias-variance decomposition of the GVIM.

\section*{Acknowledgment}
We thank Dr. Radford Neal, Professor Emeritus at the University of Toronto, for his valuable suggestions regarding this work.

%

\newpage
\section{Appendix}

\subsection{Proof of theorem 1}
\label{stat.proofs.1}
\begin{equation}
	\begin{split}
		e_{\text{orig}}(f_0) & = \mathbb{E}_{Y,X,Z}\left(\left(Y^{(a)} - f_0( X^{(a)}, Z^{(a)})\right)^{2}\right) \\
		& = \mathbb{E}_{Y,X,Z}\left(\left(Y^{(a)} - \mathbb{E}(Y^{(a)}\mid X^{(a)}, Z^{(a)})\right)^{2}\right) \\
		& = \mathbb{E}_{X}\mathbb{E}_{Z|X}\mathbb{E}_{Y_{X}|Z}\left(\left(Y^{(a)} - \mathbb{E}(Y^{(a)}\mid X^{(a)}, Z^{(a)})\right)^{2} \right)\\
		& = \mathbb{E}_{X}\mathbb{E}_{Z|X}\left(\mathbb{V}(Y|X, Z)\right) \\
		& = \sum_{k\in \{1,2,..., K\}}p_{k}\mathbb{E}_{Z|X=k}\left(\mathbb{V}(Y| X =k, Z = z)\right)
	\end{split}
	\label{orig.expand}
\end{equation}

Here, $p_k$ is the marginal probability of the $k$th treatment. From \eqref{switch} we can write,

\begin{equation}
	\begin{split}
		e_{\text{switch}}(f_0) & = \mathbb{E}_{Y,X,Z}\left(\left(Y^{(a)} - f_{0}(X^{(b)}, Z^{(a)})\right)^2\right) \\
		& = \mathbb{E}_{X^{(b)}}\mathbb{E}_{X^{(a)}}\mathbb{E}_{Z^{(a)}|X^{(a)}}\mathbb{E}_{Y^{(a)}|X^{(a)},Z^{(a)}}\left(\left( Y^{(a)}- \mathbb{E}(Y\mid \PO{X}{b} ,Z^{(a)})\right)^{2}\right)
	\end{split}
	\label{switch.expand1}
\end{equation}

First, we expand the outer most expectation on $X$ in \eqref{switch.expand1},

\begin{equation}
	e_{\text{switch}}(f_0) = \sum_{k,j\in\{1,...,K\}}\mathbb{P}(X_{(a)} = k, X_{(b)} = j)\mathbb{E}_{Z^{(a)}|X^{(a)}}\mathbb{E}_{Y^{(a)}|X^{(a)} = k, Z = z^{(a)}}\left( \left( Y^{(a)}- \mathbb{E}(Y\mid X^{(b)} = j, Z = z^{(a)})\right)^{2}\right)
	\label{switch.expand2}
\end{equation}

It is assumed that the treatment assignments are independent of each other. That is $X^{(b)}\perp X^{(a)}$. Thus, $\mathbb{P}(X_{(a)} = k, X_{(b)}= j) = \mathbb{P}(X_{(a)} = k)\mathbb{P}(X_{(b)} = j)$. Treatment $k$ and $j$ can switch in two places $a$ and $b$. Thus,
\begin{equation}
	\begin{split}
		\mathbb{P}(X_{(a)} = k)\mathbb{P}(X_{(b)} = j) & = p_{k}^{(I(X^{(a)} = k) + I(X^{(b)} = k))}p_{j}^{2 - (I(X^{(a)} = k) + I(X^{(b)} = k))} \\
		& = p_{k}^{i + i^{\prime}}p_{j}^{2 - i - i^{\prime}}
	\end{split}
\end{equation}

Assuming $i = I(X^{(a)} = k)$ and $i^{\prime} = I(X^{(b)} = j)$, where $I(.)$ is an indicator function.  Then, \eqref{switch.expand2} can be re-written as,,

\begin{equation}
	\begin{split}
		e_{\text{switch}}(f_0) = \sum_{k,j\in\{1,...,K\}}p_{k}^{i + i^{\prime}}p_{j}^{2 - i - i^{\prime}}\mathbb{E}_{Z^{(a)}|X^{(a)}}\mathbb{E}_{Y^{(a)}| X^{(a)} = k, Z^{(a)} = z}\left(\left( Y^{(a)}- \mathbb{E}(Y\mid X^{(b)} = j, Z = z^{(a)})\right)^{2}\right)
	\end{split}
	\label{switch.expand3}
\end{equation}

Since we know $X^{(a)} = k$ and $X^{(b)} = j$, we can get rid of the subscripts $a$ and $b$ (e.g., $Z^{(a)} = z$) in \eqref{switch.expand3}

\begin{equation}
	\begin{split}
		e_{\text{switch}}(f_0) & = \sum_{k,j\in\{1,...,K\}}p_{k}^{i + i^{\prime}}p_{j}^{2 - i - i^{\prime}}\mathbb{E}_{Z|X = k}\mathbb{E}_{Y|X = k, Z = z}\left(\left( Y- \mathbb{E}(Y\mid X = j, Z = z)\right)^{2}\right) \\
		& = \sum_{k,j\in\{1,...,K\}} A_{kj}
	\end{split}
	\label{switch.expand4}
\end{equation}

When $k = j$, we get two distinct terms from \eqref{switch.expand4}. They are,

\begin{equation*}
	A_{kk} = p_{k}^{2}\mathbb{E}_{Z|X= k}\mathbb{V}(Y\mid X = k,  Z = z)
\end{equation*}

and 

\begin{equation*}
	A_{jj} = p_{j}^{2}\mathbb{E}_{Z|X= j}\mathbb{V}(Y\mid X = j, Z = z)
\end{equation*}

when $k\neq j$, again \eqref{switch.expand4} produces two distinct terms,

\begin{equation}
	\begin{split}
		A_{kj} & = p_{k}p_{j}\mathbb{E}_{Z|X = k}\mathbb{E}_{Y|X = k, Z = z}\left(Y - \mathbb{E}(Y|X = j, Z = z)\right)^{2} \\
		& = p_{k}p_{j}\mathbb{E}_{Z|X = k}\mathbb{E}_{Y|X = k, Z = z}\left(Y^{2} - 2Y\mathbb{E}(Y| X = j, Z = z)+\mathbb{E}(Y|X = j, Z = z)^2\right) \\
		& = p_{k}p_{j}\mathbb{E}_{Z|X = k}\left(\mathbb{E}(Y^2|X = k, Z = z) - 2\mathbb{E}(Y|X = k, Z = z)\mathbb{E}(Y|X = j, Z = z)+\right.\\
		& \left.\mathbb{E}(Y|X = j, Z = z)^{2}\right) \\
		& = p_{k}p_{j}\mathbb{E}_{Z|X = k}\left(\mathbb{V}(Y|X = k, Z = z) + \mathbb{E}(Y|X=k, Z = z)^{2} \right.\\
		& \left. - 2\mathbb{E}(Y|X = k, Z = z)\mathbb{E}(Y|X = j, Z = z)+\mathbb{E}(Y|X = j, Z = z)^2\right) \\
		& = p_{k}p_{j}\mathbb{E}_{Z|X= k}\left(\mathbb{V}(Y|X = k, Z = z) +(\mathbb{E}(Y|X = k, Z = z) - \mathbb{E}(Y|X = j, Z = z))^2\right) \\
	\end{split}
	\label{Akj}
\end{equation}

We can apply the conditional ignorability assumption on the last line of \eqref{Akj}, 

\begin{equation}
	A_{kj}	=  p_{k}p_{j}\mathbb{E}_{Z|X = k}\left(\mathbb{V}(Y_{k}|Z = z) + \mathbb{E}_{Z\mid X=k}(\mathbb{E}(Y_k | Z) - \mathbb{E}(Y_j|Z))^2\right)
\end{equation}
Similarly,

\begin{equation*}
	A_{jk}	=  p_{k}p_{j}\mathbb{E}_{Z|X = j}\left(\mathbb{V}(Y_{j}|Z = z) + \mathbb{E}_{Z\mid X=k}(\mathbb{E}(Y_j | Z) - \mathbb{E}(Y_k|Z))^2\right)
\end{equation*}

Let's denote, $\mathbb{E}(Y_j | Z) - \mathbb{E}(Y_k|Z) = \text{CATE}_{jk}(z)$ and $\mathbb{E}(Y_k | Z) - \mathbb{E}(Y_j|Z) = \text{CATE}_{jk}(z)$ 

Incorporating these $A_{kj}$ terms in \eqref{switch.expand4} we get,

\begin{equation*}
	\begin{split}
		e_{\text{switch}}(f_0)  = & \sum_{k,j\in\{1,...,K\}}\Big(p_{k}^{2}\mathbb{E}_{Z|X= k}\mathbb{V}(Y_{k}\mid Z = z) + p_{j}^{2}\mathbb{E}_{Z|X = j}\mathbb{V}(Y_{j}\mid Z = z) + \\
		& p_{k}p_{j}\mathbb{E}_{Z|X = k}\left(\mathbb{V}(Y_{k}|Z = z) + \text{CATE}_{kj}(z)^{2}\right) + p_{k}p_{j}\mathbb{E}_{Z|X = j}\left(\mathbb{V}(Y_{j}|Z = z) + \text{CATE}_{kj}(z)^{2}\right)\Big) \\
		= & p_{1}^{2} \mathbb{E}_{Z|X= 1}\mathbb{V}(Y_{1}\mid Z = z) + p_{2}^{2}\mathbb{E}_{Z|X= 2}\mathbb{V}(Y_{2}\mid Z = z) +... \\
		& + p_{1}p_{2}\mathbb{E}_{Z|X=1}(\mathbb{V}(Y_{1}|Z = z)+\text{CATE}_{12}(z)^{2}) +  p_{1}p_{2}\mathbb{E}_{Z|X=2}(\mathbb{V}(Y_{2}|Z = z)+\text{CATE}_{12}(z)^{2}) ... \\
		= & p_{1}(p_{1}+p_{2}+...+p_{K})\mathbb{E}_{Z|X = 1}\mathbb{V}(Y_{1}\mid Z = z) + p_{2}(p_{1}+p_{2}+...+p_{K})\mathbb{E}_{Z|X = 2}\mathbb{V}(Y_{2}\mid Z = z) + ... \\ 
		&+ p_{1}p_{2}\mathbb{E}_{Z|X=1}\text{CATE}_{12}(z)^{2} + p_{1}p_{2}\mathbb{E}_{Z|X=2}\text{CATE}_{12}(z)^{2}+.... \\
		= & p_{1}\mathbb{E}_{Z|X = 1}\mathbb{V}(Y_{1}\mid Z = z) + p_{2}\mathbb{E}_{Z|X = 2}\mathbb{V}(Y_{2}\mid Z = z) + ... +\sum_{k,j\in\{1,...K\}}p_{k}p_{j}\mathbb{E}_{Z|X=k}\text{CATE}_{jk}(z)^{2} \\
		= & e_{\text{orig}}(f_0) + \sum_{k,j\in\{1,...K\}}p_{k}p_{j}\mathbb{E}_{Z|X=k}\text{CATE}_{jk}(z)^{2} \\
		e_{\text{switch}}(f_0) =&  e_{\text{orig}}(f_0) + \sum_{k\neq j}p_{k}p_{j}\mathbb{E}_{Z|X=k}\text{CATE}_{kj}(z)^{2} \\
	\end{split}
\end{equation*}

That is, from \eqref{vim} we get,

\begin{equation}
	\begin{split}
		GVIM_{X}(f_0) = & e_{\text{switch}}(f_0)  - e_{\text{orig}}(f_0) \\
		= &  \sum_{k\neq j}p_{k}p_{j}\mathbb{E}_{Z|X=k}\text{CATE}_{kj}(z)^{2}
		\label{vim.12}
	\end{split}
\end{equation}

\subsection{Proof of theorem 2}
\label{stat.proofs.2}
\begin{equation}
	\begin{split}
		e_{\text{orig}}(f_0) & = \mathbb{E}_{Y,X,Z}\left(\left(Y^{(a)} - f_0( X^{(a)}, Z^{(a)})\right)^{2}\right) \\
		& = \mathbb{E}_{Y,X,Z}\left(\left(\PO{Y}{a} - \mathbb{E}(\PO{Y}{a}\mid \PO{X}{a}, \PO{Z}{a})\right)^{2}\right) \\
		& =\mathbb{E}_{X}\mathbb{E}_{Z|X} \mathbb{V}(Y\mid \PO{X}{a}, \PO{Z}{a})
	\end{split}
	\label{orig.expand.cont}
\end{equation}

Recall the general form of loss in \eqref{switch.expand1} after switching $\PO{X}{a}$ and $\PO{X}{b}$ was,

\begin{equation}
	\begin{split}
		e_{\text{switch}}(f_0) & = \mathbb{E}_{Y,X,Z}\left(\left(Y^{(a)} - f_{0}(X^{(b)}, Z^{(a)})\right)^2\right) \\
		& = \mathbb{E}_{X^{(b)}}\mathbb{E}_{X^{(a)}}\mathbb{E}_{Z^{(a)}|X^{(a)}}\mathbb{E}_{Y^{(a)}|X^{(a)}, Z^{(a)}}\left(\left( Y^{(a)} - \mathbb{E}(Y^{(a)}\mid X^{(b)}, Z^{(a)})\right)^{2}\right) \\
		& = \mathbb{E}_{X^{(b)}}\mathbb{E}_{X^{(a)}}\mathbb{E}_{Z^{(a)}|X^{(a)}}\mathbb{E}_{Y|X^{(a)}, Z^{(a)}}\left(\left( Y^{(a)} - \mathbb{E}(Y\mid X^{(b)}, Z^{(a)})\right)^{2}\right) \\
	\end{split}
	\label{cont.switch}
\end{equation}

Expanding the inner-most expectation of \eqref{cont.switch} we get,

\begin{equation}
	\begin{split}
		&	\mathbb{E}_{Y|X^{(a)}, Z^{(a)}}\left(\left( Y^{(a)} - \mathbb{E}(Y\mid X^{(b)}, Z^{(a)})\right)^{2}\right) \\
		= & 	\mathbb{E}_{Y|X^{(a)},Z^{(a)}}\left((Y^{(a)})^2- 2 Y^{(a)}  \mathbb{E}(Y\mid X^{(b)}, Z^{(a)}) +  \left(\mathbb{E}(Y\mid X^{(b)}, Z^{(a)})\right)^2\right) \\
		= & \mathbb{E}_{Y|X^{(a)},Z^{(a)}} \left( (Y^{(a)})^2\right) - 2\mathbb{E}_{Y|X^{(a)},Z^{(a)}}(Y^{(a)})\mathbb{E}(Y|X^{(b)}, Z^{(a)}) + \left(\mathbb{E}(Y|X^{(b)},Z^{(a)})\right)^2 \\
		= & \mathbb{E}\left( \left(Y^{(a)}|X^{(a)}, Z^{(a)}\right)^2\right) - 2\mathbb{E}(Y|X^{(a)}, Z^{(a)})\mathbb{E}(Y|X^{(b)}, Z^{(a)}) +\left( \mathbb{E}(Y|X^{(b)},Z^{(a)})\right)^2 \\
		= & \mathbb{V}(Y \mid  X^{(a)}, Z^{(a)}) + \left(\mathbb{E}(Y\mid X^{(a)}, Z^{(a)}\right)^2 - 2\mathbb{E}(Y\mid X^{(a)}, Z^{(a)})\mathbb{E}(Y\mid X^{(b)}, Z^{(a)}) + \left(\mathbb{E}(Y\mid  X^{(b)}, Z^{(a)}\right)^2 \\
		= & \mathbb{V}(Y \mid  X^{(a)}, Z^{(a)})  + \left( \mathbb{E}(Y\mid X^{(a)}, Z^{(a)}) - \mathbb{E}(Y \mid X^{(b)}, Z^{(a)}) \right)^2\\
	\end{split}
	\label{switch.expand.cont}
\end{equation}

From using the conditional ignorability assumption we know that $\mathbb{E}(Y | X = x, Z) = \mathbb{E}(Y_{x}|Z)$

\begin{equation}
	\begin{split}
		& \mathbb{V}(Y \mid  X^{(a)}, Z^{(a)})  + \left( \mathbb{E}(Y\mid X^{(a)}, Z^{(a)}) - \mathbb{E}(Y \mid X^{(b)}, Z^{(a)})\right)^2 \\
		= & \mathbb{V}(Y_{X^{(a)}} \mid  Z^{(a)}) +  \left( \mathbb{E}(Y_{X^{a}}\mid  Z^{(a)}) - \mathbb{E}(Y_{X^{b}} \mid Z^{(a)})\right)^2\\
		= & \mathbb{V}(Y_{X^{(a)}}\mid Z) + \left(\mathbb{E}\left(Y_{X^{(a)}} - Y_{X^{(b)}} | Z\right)\right)^2
	\end{split}
	\label{switch.expand.cont2}
\end{equation}

The final line is obtained by applying the assumption of conditional ignorability. 

Thus inserting \eqref{switch.expand.cont2} in \eqref{switch.expand1} we get,

\begin{equation}
	\begin{split}
		e_{\text{switch}}(f_0) & = \mathbb{E}_{Y,X,Z}\left(\left(Y^{(a)} - f_{0}(X^{(b)}, Z^{(a)})\right)^2\right) \\
		& = \mathbb{E}_{X^{(b)}}\mathbb{E}_{X^{(a)}}\mathbb{E}_{Z^{(a)}|X^{(a)}}\mathbb{E}_{Y^{(a)}|X^{(a)},Z^{(a)}}\left(\left( Y^{(a)} - \mathbb{E}(Y\mid X^{(b)}, Z^{(a)})\right)^{2}\right) \\
		& = \mathbb{E}_{X}\mathbb{E}_{Z|X} \left(\mathbb{V}(Y_{X^{(a)}}\mid Z) \right) + \mathbb{E}_{X^{(b)}}\mathbb{E}_{X^{(a)}}\mathbb{E}_{Z^{(a)}|X^{(a)}} \left(\mathbb{E}\left(Y_{X^{(a)}} - Y_{X^{(b)}} | Z\right)\right)^2\\
		& = e_{\text{orig}}(f_0)   + \mathbb{E}_{X^{(b)}}\mathbb{E}_{X^{(a)}}\mathbb{E}_{Z|X^{(a)}} \left(\mathbb{E}\left(Y_{X^{(a)}} - Y_{X^{(b)}} | Z\right)\right)^2
	\end{split}
	\label{VIM.cont1}
\end{equation}

\subsection{\%Bias table for the remaining variables }
\begin{table}[H]
	\centering
	\caption{Table for the \%Bias in estimated GVIM for all the important predictors }
	\label{tab.A2}
	\begin{tabular}{rrrrrrrrr}
		\hline
		& & & \multicolumn{2}{c}{oracle} &  \multicolumn{2}{c}{XGBoost} &  \multicolumn{2}{c}{GAM} \\
		\hline
		& Training Size & True & $\frac{\hat{e}_{orig}}{e_{\text{orig}}}$ & \%Bias & $\frac{\hat{e}_{orig}}{e_{\text{orig}}}$ & \%Bias & $\frac{\hat{e}_{orig}}{e_{\text{orig}}}$ & \%Bias \\ 
		\hline
		$C_2$ & 50 & 3.11 & 1.27 & -1.05 & 36.81 & -83.75 & 35.48 & -4.70 \\ 
   & 500 & 3.11 & 1.02 & 0.45 & 9.97 & -27.44 & 6.19 & 4.36 \\ 
   & 5000 & 3.11 & 1.01 & 0.26 & 3.04 & -6.23 & 5.47 & 1.71 \\ 
   & 50000 & 3.11 & 1.00 & 0.01 & 1.46 & -1.07 & 5.35 & 1.26 \\ 
  $U_2$ & 50 & 3.10 & 1.27 & 1.50 & 36.81 & -102.44 & 35.48 & -104.01 \\ 
   & 500 & 3.10 & 1.02 & 0.70 & 9.97 & -92.92 & 6.19 & -99.97 \\ 
   & 5000 & 3.10 & 1.01 & 0.24 & 3.04 & -56.31 & 5.47 & -100.00 \\ 
   & 50000 & 3.10 & 1.00 & 0.06 & 1.46 & -20.33 & 5.35 & -100.00 \\ 
		\hline
	\end{tabular}    
\end{table}

\subsection{Boxplots of estimated VIM for the remaining variables}

\begin{figure}[H]
	\centering
	\includegraphics[scale = 0.6]{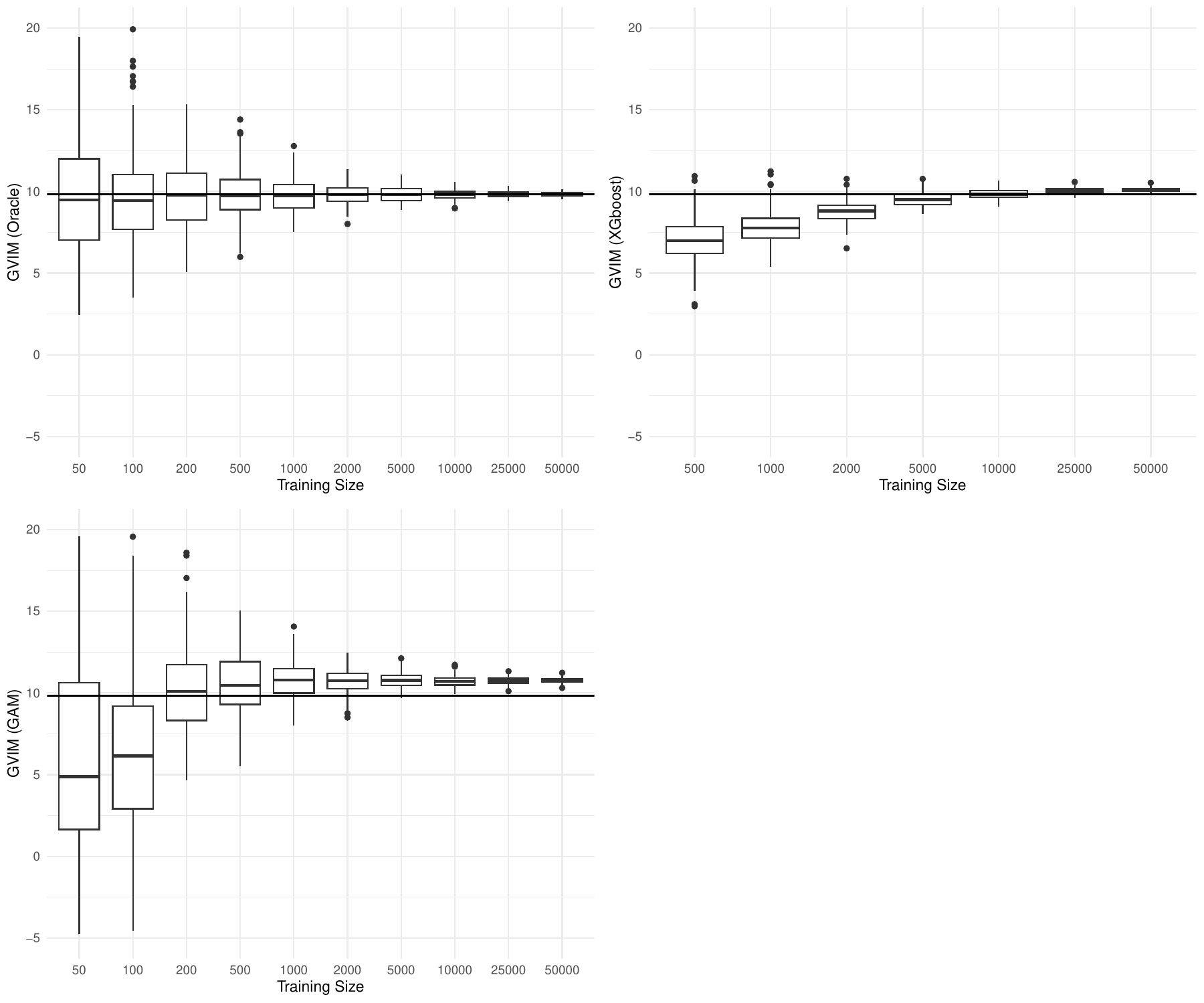}
	\caption{Estimated GVIM for the predictor $X_3$ }
	\label{Fig7A}
\end{figure}

\begin{figure}[H]
	\centering
	\includegraphics[scale = 0.6]{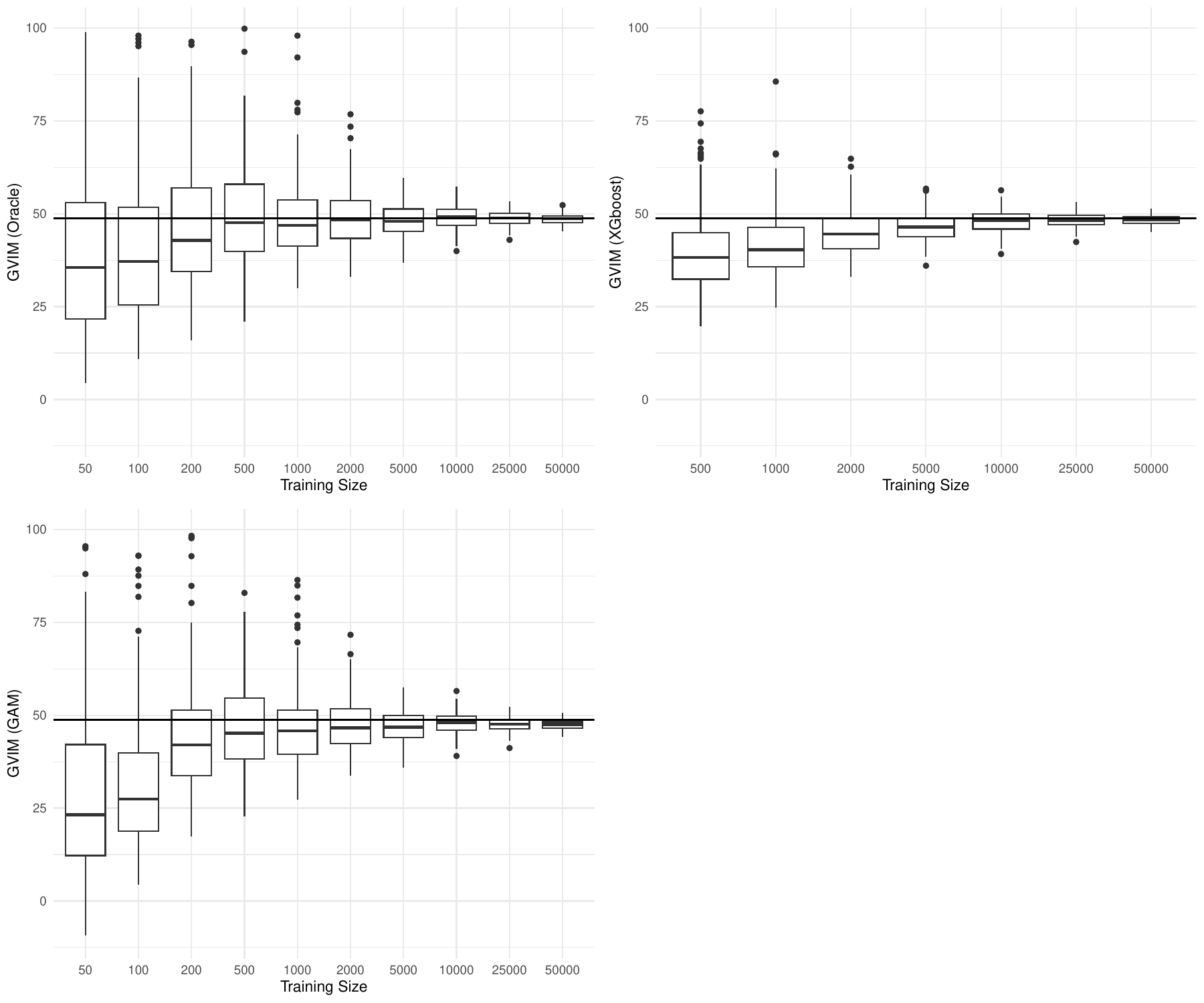}
	\caption{Estimated GVIM for the predictor $X_4$ }
	\label{Fig5}
\end{figure}

\begin{figure}[H]
	\centering
	\includegraphics[scale = 0.6]{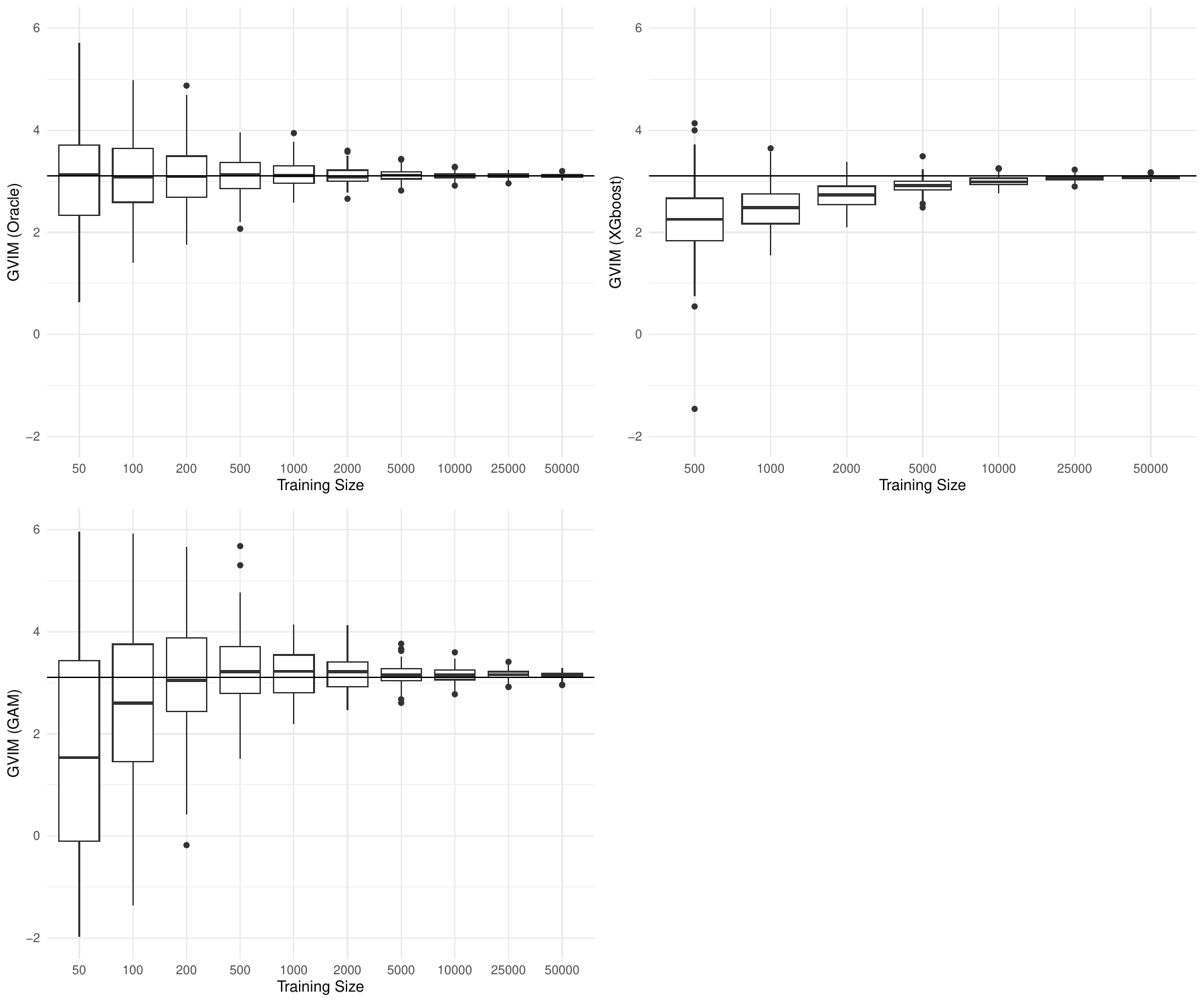}
	\caption{Estimated GVIM for the predictor $C_2$ }
	\label{Fig6A}
\end{figure}

\begin{figure}[H]
	\centering
	\includegraphics[scale = 0.6]{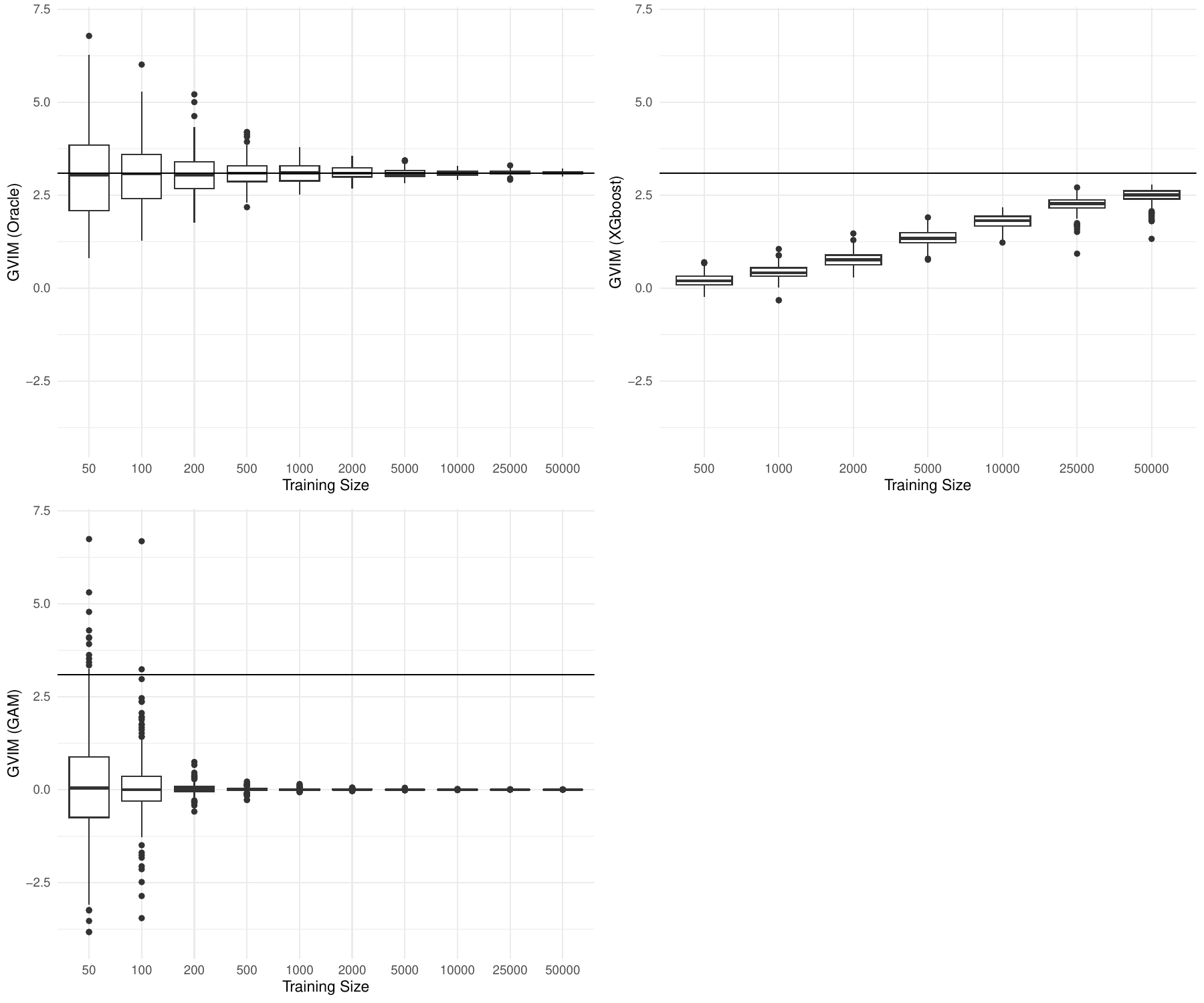}
	\caption{Estimated GVIM for the predictor $U_1$ }
	\label{Fig8A}
\end{figure}

\begin{figure}[H]
	\centering
	\includegraphics[scale = 0.6]{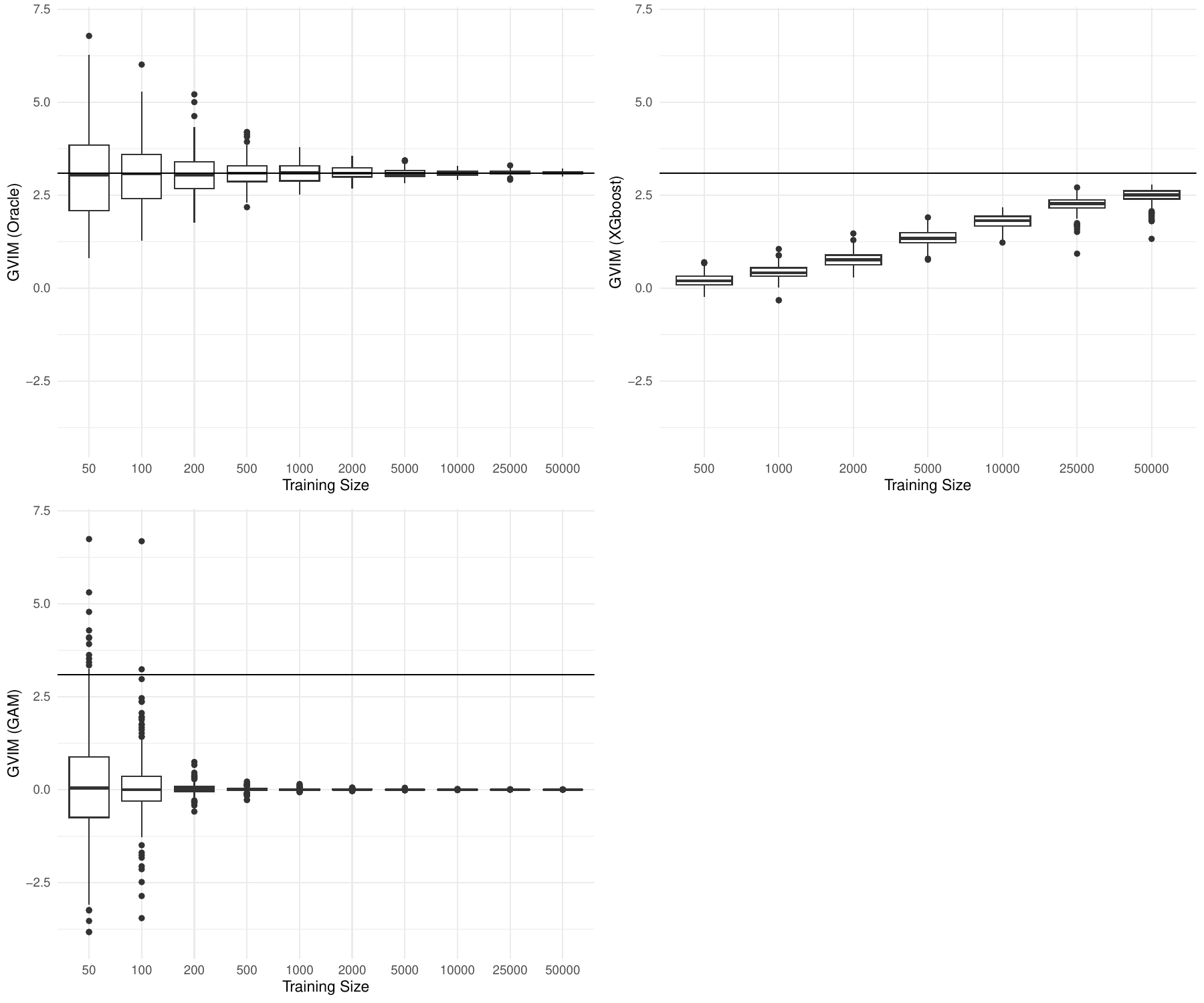}
	\caption{Estimated GVIM for the predictor $U_2$ }
	\label{Fig9A}
\end{figure}

\bibliography{kak_thesis}{}

\begin{thebibliography}{27}
\expandafter\ifx\csname natexlab\endcsname\relax\def\natexlab#1{#1}\fi
\expandafter\ifx\csname url\endcsname\relax
  \def\url#1{\texttt{#1}}\fi
\expandafter\ifx\csname urlprefix\endcsname\relax\def\urlprefix{URL }\fi

\bibitem[{Molnar(2020)}]{molnar2020interpretable}
Molnar, C.
\newblock \emph{Interpretable machine learning} (Lulu. com, 2020).

\bibitem[{Breiman(2001)}]{breiman2001random}
Breiman, L.
\newblock Random forests.
\newblock \emph{Machine learning} \textbf{45}, 5--32 (2001).

\bibitem[{Ribeiro \emph{et~al.}(2016)Ribeiro, Singh \&
  Guestrin}]{ribeiro2016should}
Ribeiro, M.~T., Singh, S. \& Guestrin, C.
\newblock " Why should i trust you?" Explaining the predictions of any
  classifier.
\newblock In \emph{Proceedings of the 22nd ACM SIGKDD international conference
  on knowledge discovery and data mining}, 1135--1144 (2016).

\bibitem[{Lundberg \& Lee(2017)}]{lundberg2017unified}
Lundberg, S.~M. \& Lee, S.-I.
\newblock A unified approach to interpreting model predictions.
\newblock In \emph{Proceedings of the 31st international conference on neural
  information processing systems}, 4768--4777 (2017).

\bibitem[{Weller \emph{et~al.}(2021)}]{weller2021predicting}
Weller, O., Sagers, L., Hanson, C., Barnes, M., Snell, Q. \& Tass, E.~S.
\newblock Predicting suicidal thoughts and behavior among adolescents using the
  risk and protective factor framework: A large-scale machine learning
  approach.
\newblock \emph{Plos one} \textbf{16}, e0258535 (2021).

\bibitem[{Gregorutti \emph{et~al.}(2017)Gregorutti, Michel \&
  Saint-Pierre}]{gregorutti2017correlation}
Gregorutti, B., Michel, B. \& Saint-Pierre, P.
\newblock Correlation and variable importance in random forests.
\newblock \emph{Statistics and Computing} \textbf{27}, 659--678 (2017).

\bibitem[{Fisher \emph{et~al.}(2019{\natexlab{a}})Fisher, Rudin \&
  Dominici}]{fisher2019all}
Fisher, A., Rudin, C. \& Dominici, F.
\newblock All Models are Wrong, but Many are Useful: Learning a Variable's
  Importance by Studying an Entire Class of Prediction Models Simultaneously.
\newblock \emph{J. Mach. Learn. Res.} \textbf{20}, 1--81 (2019{\natexlab{a}}).

\bibitem[{Hooker \emph{et~al.}(2021)Hooker, Mentch \&
  Zhou}]{hooker2021unrestricted}
Hooker, G., Mentch, L. \& Zhou, S.
\newblock Unrestricted permutation forces extrapolation: variable importance
  requires at least one more model, or there is no free variable importance.
\newblock \emph{Statistics and Computing} \textbf{31}, 1--16 (2021).

\bibitem[{Strobl \emph{et~al.}(2007)Strobl, Boulesteix, Zeileis \&
  Hothorn}]{strobl2007bias}
Strobl, C., Boulesteix, A.-L., Zeileis, A. \& Hothorn, T.
\newblock Bias in random forest variable importance measures: Illustrations,
  sources and a solution.
\newblock \emph{BMC bioinformatics} \textbf{8}, 1--21 (2007).

\bibitem[{Strobl \emph{et~al.}(2008)Strobl, Boulesteix, Kneib, Augustin \&
  Zeileis}]{strobl2008conditional}
Strobl, C., Boulesteix, A.-L., Kneib, T., Augustin, T. \& Zeileis, A.
\newblock Conditional variable importance for random forests.
\newblock \emph{BMC bioinformatics} \textbf{9}, 1--11 (2008).

\bibitem[{Kaneko(2022)}]{kaneko2022cross}
Kaneko, H.
\newblock Cross-validated permutation feature importance considering
  correlation between features.
\newblock \emph{Analytical Science Advances} \textbf{3}, 278--287 (2022).

\bibitem[{Diaz \emph{et~al.}(2015)Diaz, Hubbard, Decker \&
  Cohen}]{diaz2015variable}
Diaz, I., Hubbard, A., Decker, A. \& Cohen, M.
\newblock Variable importance and prediction methods for longitudinal problems
  with missing variables.
\newblock \emph{PloS one} \textbf{10} (2015).

\bibitem[{Van~der Laan \& Rose(2011)}]{van2011targeted}
Van~der Laan, M.~J. \& Rose, S.
\newblock \emph{Targeted learning: causal inference for observational and
  experimental data} (Springer Science \& Business Media, 2011).

\bibitem[{Buhlmann \emph{et~al.}(2002)Buhlmann, Yu
  \emph{et~al.}}]{buhlmann2002analyzing}
Buhlmann, P., Yu, B. \emph{et~al.}
\newblock Analyzing bagging.
\newblock \emph{Annals of statistics} \textbf{30}, 927--961 (2002).

\bibitem[{Jiang \& Owen(2002)}]{jiang2002quasi}
Jiang, T. \& Owen, A.~B.
\newblock Quasi-regression for visualization and interpretation of black box
  functions (2002).

\bibitem[{Zhao \& Hastie(2019)}]{zhao2019causal}
Zhao, Q. \& Hastie, T.
\newblock Causal interpretations of black-box models.
\newblock \emph{Journal of Business \& Economic Statistics} 1--10 (2019).

\bibitem[{Van~der Laan \& Rose(2018)}]{van2018targeted}
Van~der Laan, M.~J. \& Rose, S.
\newblock \emph{Targeted learning in data science} (Springer, 2018).

\bibitem[{Gruber \& van~der Laan(2010)}]{gruber2010application}
Gruber, S. \& van~der Laan, M.~J.
\newblock An application of collaborative targeted maximum likelihood
  estimation in causal inference and genomics.
\newblock \emph{The International Journal of Biostatistics} \textbf{6} (2010).

\bibitem[{Rinaldo \emph{et~al.}(2019)Rinaldo, Wasserman \&
  G’Sell}]{rinaldo2019bootstrapping}
Rinaldo, A., Wasserman, L. \& G’Sell, M.
\newblock Bootstrapping and sample splitting for high-dimensional,
  assumption-lean inference.
\newblock \emph{The Annals of Statistics} \textbf{47}, 3438--3469 (2019).

\bibitem[{Fisher \emph{et~al.}(2019{\natexlab{b}})Fisher, Smith \&
  Walsh}]{fisher2019machine}
Fisher, C.~K., Smith, A.~M. \& Walsh, J.~R.
\newblock Machine learning for comprehensive forecasting of Alzheimer’s
  Disease progression.
\newblock \emph{Scientific reports} \textbf{9}, 1--14 (2019{\natexlab{b}}).

\bibitem[{Gregorutti \emph{et~al.}(2015)Gregorutti, Michel \&
  Saint-Pierre}]{gregorutti2015grouped}
Gregorutti, B., Michel, B. \& Saint-Pierre, P.
\newblock Grouped variable importance with random forests and application to
  multiple functional data analysis.
\newblock \emph{Computational Statistics \& Data Analysis} \textbf{90}, 15--35
  (2015).

\bibitem[{Verdinelli \& Wasserman(2023)}]{verdinelli2023feature}
Verdinelli, I. \& Wasserman, L.
\newblock Feature Importance: A Closer Look at Shapley Values and LOCO.
\newblock \emph{arXiv preprint arXiv:2303.05981}  (2023).

\bibitem[{Efron \& Tibshirani(1997)}]{efron1997improvements}
Efron, B. \& Tibshirani, R.
\newblock Improvements on cross-validation: the 632+ bootstrap method.
\newblock \emph{Journal of the American Statistical Association} \textbf{92},
  548--560 (1997).

\bibitem[{Tibshirani \& Efron(1993)}]{tibshirani1993introduction}
Tibshirani, R.~J. \& Efron, B.
\newblock An introduction to the bootstrap.
\newblock \emph{Monographs on statistics and applied probability} \textbf{57},
  1--436 (1993).

\bibitem[{Breiman(1996)}]{breiman1996bagging}
Breiman, L.
\newblock Bagging predictors.
\newblock \emph{Machine learning} \textbf{24}, 123--140 (1996).

\bibitem[{Friedman(1991)}]{friedman1991multivariate}
Friedman, J.~H.
\newblock Multivariate adaptive regression splines.
\newblock \emph{The annals of statistics} \textbf{19}, 1--67 (1991).

\bibitem[{Chen \& Guestrin(2016)}]{chen2016xgboost}
Chen, T. \& Guestrin, C.
\newblock Xgboost: A scalable tree boosting system.
\newblock In \emph{Proceedings of the 22nd acm sigkdd international conference
  on knowledge discovery and data mining}, 785--794 (2016).

\end{thebibliography}
\bibliographystyle{nature}

\end{document}